\documentclass[apjl, iop]{emulateapj}

\newcommand \Hbeta {\ifmmode {\rm H}\beta \else H$\beta$\fi}
\newcommand \ha    {\ifmmode {\rm H}\alpha \else H$\alpha$\fi}
\newcommand \hb    {\ifmmode {\rm H}\beta \else H$\beta$\fi}
\newcommand \hg    {\ifmmode {\rm H}\gamma \else H$\gamma$\fi}
\newcommand \hd    {\ifmmode {\rm H}\delta \else H$\delta$\fi}
\newcommand  \mgii  {\ifmmode {\rm Mg}{\textsc{ii}} \else Mg\,{\sc ii}\fi}
\newcommand  \MGII  {\ifmmode {\rm Mg}\,{\sc ii}\,\lambda2798 \else Mg\,{\sc ii}\,$\lambda2798$\fi}
\newcommand  \siiv  {\ifmmode {\rm Si}\, {\sc iv}\ \else Si\,{\sc iv}\fi}
\newcommand  \SIIV  {\ifmmode {\rm Si}\,{\sc iv}\,\lambda1399 \else Si\,{\sc iv}\,$\lambda1399$\fi}
\newcommand  \civ  {\ifmmode {\rm C}\, {\sc iv}\ \else C\,{\sc iv}\fi}
\newcommand  \CIV  {\ifmmode {\rm C}\,{\sc iv}\,\lambda1549 \else C\,{\sc iv}\,$\lambda1549$\fi}
\newcommand  \NV  {\ifmmode {\rm N}\,{\sc v}\,\lambda1240 \else N\,{\sc v}\,$\lambda1240$\fi}
\newcommand  \nv  {\ifmmode {\rm N}\,{\sc v}\ \else N\,{\sc v}\fi}
\newcommand  \pv  {\ifmmode {\rm P}\,{\sc v}\ \else P\,{\sc v}\fi}
\newcommand  \LyA  {\ifmmode {\rm Lyman}\,{\sc $\alpha$}\,\lambda1216 \else Lyman\,{\sc $\alpha$}\,$\lambda1216$\fi}
\newcommand  \feii     {Fe\,{\sc ii}}

\newcommand  \aliii  {\ifmmode {\rm Al}{\textsc{iii}} \else Al\,{\sc iii}\fi}
\newcommand  \ALIII  {\ifmmode {\rm Al}\,{\sc iii}\,\lambda1857 \else Al\,{\sc iii}\,$\lambda1857$\fi}

\newcommand  \CIII  {\ifmmode {\rm C}\,{\sc iii]}\,\lambda1909 \else C\,{\sc iii]}\,$\lambda1909$\fi}
\newcommand  \oi    {\ifmmode \left[{\rm O}\,{\textsc i}\right] \else [O\,{\sc i}]\fi}
\newcommand  \OI    {\ifmmode \left[{\rm O}\,{\textsc i}\right]\,\lambda6300 \else [O\,{\sc i}]$\,\lambda6300$ \fi}
\newcommand  \hei     {He\,{\sc i*}}
\newcommand  \oii   {\ifmmode \left[{\rm O}\,{\textsc ii}\right] \else [O\,{\sc ii}]\fi}
\newcommand  \OII   {\ifmmode \left[{\rm O}\,{\textsc ii}\right]\,\lambda3727 \else [O\,{\sc ii}]\,$\lambda3727$ \fi}
\newcommand  \oiii  {\ifmmode \left[{\rm O}\,{\textsc iii}\right] \else [O\,{\sc iii}]\fi}
\newcommand  \OIII  {\ifmmode \left[{\rm O}\,{\textsc iii}\right]\,\lambda5007 \else [O\,{\sc iii}]\,$\lambda5007$\fi}
\newcommand  \ovi    {\ifmmode \left[{\rm O}\,{\textsc vi}\right] \else O\,{\sc vi}\fi}
\newcommand  \neiii   {\ifmmode \left[{\rm Ne}\,{\textsc iii}\right] \else [Ne\,{\sc iii}]\fi}
\newcommand  \nev   {\ifmmode \left[{\rm Ne}\,{\textsc v}\right] \else [Ne\,{\sc v}]\fi}

\newcommand{\kms}{\, {\rm km\, s}^{-1}}
\newcommand{\acms}{\, {\rm cm\, s}^{-2}}

\usepackage{ulem}

\shorttitle{LoBAL acceleration}
\shortauthors{Yi et al.}

\begin{document}


\title{Quasar winds  caught on acceleration and deceleration}

\author{Weimin Yi\altaffilmark{1}, P. B. Hall\altaffilmark{2},  Zunlin Yuan\altaffilmark{3}, W.~N. Brandt\altaffilmark{4,5},  D.~P. Schneider\altaffilmark{4,5}, Zhicheng He\altaffilmark{6}, Jin-Ming Bai\altaffilmark{1}, Xue-Bing Wu\altaffilmark{7,8}    }

 
\altaffiltext{1}{Yunnan Observatories, Kunming, 650216, China}
\altaffiltext{2}{Department of Physics and Astronomy, York University, Toronto, ON, M3J 1P3, Canada} 
\altaffiltext{3}{Department of Physics, School of Physics and Electronics, Hunan Normal University, Changsha 410081, People's Republic of China}
\altaffiltext{4}{Department of Astronomy \& Astrophysics, The Pennsylvania State University, 525 Davey Lab, University Park, PA 16802, USA}
\altaffiltext{5}{Institute for Gravitation and the Cosmos, The Pennsylvania State University, University Park, PA 16802, USA}
\altaffiltext{6}{CAS Key Laboratory for Research in Galaxies and Cosmology, Department of Astronomy, University of Science and Technology \\ of China, Hefei, Anhui 230026, China}
\altaffiltext{7}{Kavli Institute for Astronomy and Astrophysics, Peking University, Beijing 100871, People’s Republic of China }
\altaffiltext{8}{Department of Astronomy, Peking University, Yi He Yuan Lu 5, Hai Dian District, Beijing 100871, People’s Republic of China }

\begin{abstract}
We present an observational  study of  wind acceleration  based on  four low-ionization broad absorption line (LoBAL) quasars (J0136, J1238, J1259, J1344). J0136  and J1344 (group-1)  are radio quiet  and show large  BAL-velocity  shifts as opposed to  stable  line-locking associated  absorption lines (AALs). Notably,  J1344  displays  a linear relation between BAL-velocity  shift and time interval  over three  consecutive   epochs,  characteristic of compelling evidence for BAL acceleration.  J1238 and J1259 (group 2) exhibit small BAL-velocity shifts along with steep-spectrum, weak radio emission at 3.0 and 1.4 GHz.  All four quasars have  spectral energy distributions (SEDs) with a peak  at $\lambda_{\rm rest }\sim10~\mu$m,  suggesting  a  link  between the  BAL acceleration  and hot dust emission.  The group-2 quasars are redder  than group-1 quasars and have a steeper  rise  at $1<\lambda_{\rm rest }<3~\mu$m in their SEDs.  All but J1238 exhibit a steep rise followed by a plateau-like time evolution in BAL-velocity shift. 
Our investigations, combined with previous studies of BAL acceleration,  indicate  that   (1) the BAL-ISM  coupling process  is  one of the major avenues for the origin of quasar reddening and patchy obscuration, (2) AAL  outflows  are ubiquitous and  likely signify  large-scale  remnants of  BAL winds coupled to interstellar medium (ISM),  and (3) wind deceleration that is closely linked to the BAL-ISM coupling process may produce weak  radio emission in otherwise radio-quiet quasars. 
\end{abstract}


\section{Introduction}

Broad absorption line (BAL; \citealt{Weymann91}) features imprinted on quasar spectra are unambiguous evidence for intrinsic outflows, whose kinetic power could be sufficient for triggering active galactic nuclear (AGN) feedback and hence control the growth of supermassive black holes (SMBHs; \citealp{Fabian12}).  BAL winds are believed to be launched from the accretions disks that have a typical  size  of $\sim$0.01 pc if driven by a  SMBH with $\sim10^9 M_\odot$ (\citealp{Murray95,Proga00}). Observationally, however, many studies suggest that the vast majority of BAL winds are likely to be located at a range of  $\sim$1--1000 pc from their SMBHs (e.g., \citealp{Capellupo11,McGraw17,Arav18,HeZ19}).  Therefore,  probing   inner physics of BAL winds, such as the acceleration mechanisms and the impact on ISM, has become a topic of increasing interest.

BAL quasars are divided into two major classes, namely high-ionization BAL  (HiBAL) and low-ionization BAL (LoBAL) quasars (e.g., \citealp{Weymann91,Trump06}).  A bona fide BAL is characterized by an absorption trough, whose width is broader than $2000\kms$ under  90\% of the continuum level with a minimum line-of-sight (LOS) velocity of $>3000\kms$ (\citealt{Weymann91}). To form such remarkable absorption features, acceleration of BAL winds must somehow play a role during the lifetime  of BAL quasars. \citet{Grier16} conducted the first systematic investigation of BAL acceleration, from which they found only 2 out of 140 quasars showing solid evidence of BAL acceleration. Such a low incidence ($\sim$1.43\%) of BAL acceleration is likely underestimated since their work is based on the search for monolithic velocity shifts across the entire BAL trough over multiple spectroscopic epochs. Nevertheless, one would expect from a sample of  $\sim$100,000 BAL quasars  (\citealt{Lyke20}) that a fairly large number of objects could be caught on BAL acceleration even with  an incidence of  $\sim$1.43\%.

  \citet{Hall07a} reported one of the first cases of a quasar exhibiting  BAL-acceleration signatures, whose average acceleration magnitude is  consistent with that estimated from a longer-time sampling interval in a subsequent  study (\citealp{Byun22}), although more data are needed before drawing a firm conclusion of BAL acceleration.  This point is particularly true when noticing the ``jerk'' phenomena of BAL acceleration; that is, the change in acceleration  with time  (e.g., \citealp{Capellupo11,Filizak13,Rogerson16,Grier16}).  For example, \citet{Rogerson16} identified a quasar showing BAL emergence at two widely separated velocities. Specifically, trough A showed a fluctuating velocity centroid due to trough profile variability; however, trough B showed an increasing outflow velocity over three epochs, which if confirmed to be acceleration would represent rest-frame 16 and 55 cm~s$^{-2}$ between the 1st \& 2nd and 2nd \& 3rd epochs, respectively.  Simple radiative acceleration models can match such values at small radii (see their section 4.1.2), but would have to invoke strong ionizing continuum variability to explain such large changes in acceleration. 
Similarly, \citet{Aromal21} reported a quasar having two distinct BAL troughs, one of which is consistent with BAL acceleration along with variability in rest equivalent width (REW) while the other exhibits complex variations in profile.  Recently, \citet{XuX20} argued for the largest BAL acceleration detected in a quasar, despite with only two sampling-epoch spectra. Perhaps a more ambiguous example for acceleration is  NGC 3783, a local AGN that has been extensively studied from nearly two decades of high spectral resolution observations, for which \citet{Gabel03} proposed an explanation of radial deceleration;  however, this argument appears to be increasingly uncertain in later studies (e.g., \citealp{Scott14,Kriss19}).

Another major difficulty for the analysis of BAL phenomena is the prevalence of BAL-profile variability among BAL quasars.  As introduced above, time variability  has become a widely used tool for probing the formation, acceleration, and evolution of BAL winds,  due primarily to the fact that the vast majority of quasars and their host galaxies cannot be spatially resolved by current instruments.  Generally,  well-separated, multi-epoch spectroscopy is a powerful technique  for identifying genuine BAL acceleration cases and for placing  valuable constraints on the BAL physics, lifetime, and location (e.g., \citealp{Aromal21,Yi22}).  In reality, however, BAL acceleration and BAL-profile  variability may  occur simultaneously, making a firm identification of BAL acceleration  difficult or sometimes impossible,  even with the aid of long-term, high-quality, and high-cadence sampling spectra (e.g., \citealp{Kriss19,Yi21,Byun22}). This situation  can be easily understood given   the complex nature of  BAL winds typically  characterized by many subflows that are independent from each other and vary stochastically.  Regarding the analysis of BAL acceleration,  another three facts are needed to be taken into account. First,  the  majority of BAL quasars have not been spectroscopically observed  more than three  times, rendering any attempts at identifying genuine acceleration difficult. Secondly,  BALs may be  transient  during the spectroscopic monitoring time.  Thirdly, the decomposition of a complex, seemingly single  BAL trough is notoriously difficult due to  blending and/or  partial covering along our LOS.   Nevertheless, genuine BAL acceleration or deceleration events may provide unique  insights  into the origin of quasar reddening and weak radio emission, an interesting yet open question invoked by recent studies  (e.g., \citealt{Klindt19,Rivera21}).

In this work, we conduct a systematic analysis of BAL acceleration from  the LoBAL variability sample in \citet{Yi19a} and the HiBAL acceleration  sample in \citet{Grier16}, from which we identified 4   BAL-acceleration quasars.  The first two (and in one case four) epochs of spectroscopic observations of the four quasars were performed  in various phases of the Sloan Digital Sky Survey (SDSS; \citealp{York00, Eisenstein11,  Blanton17}) and were accessed from the SDSS public database.  Additional observations were obtained with a variety of other facilities (see Section \ref{new_obs}). The new data were designed to systematically investigate LoBAL acceleration and to bridge the gap  over HiBAL acceleration.  
We emphasize   that (1) all time intervals are in the quasar rest frame; (2) the term ``acceleration'' refers to either actual acceleration or deceleration unless stated otherwise; (3)  velocity or kinematic shift, by definition,  is a relative quantity and does not depend on the accuracy of systemic redshift;  and (4) zero velocity is converted by the shorter-wavelength component of each doublet at  systemic redshift  throughout this work.

 \begin{table*}
 \center
\caption{ Spectroscopic observations of the four BAL-acceleration quasar candidates   }
\flushbottom
\begin{tabular}{|c|c|c|c|c|c|c|}
\hline
Name &Instrument& Date&MJD&Exposure Time & $\lambda$ Coverage&Resolution \\
(Abbreviation) &&(MM-DD-YYYY)&&(s)&(\AA)&($\lambda/\Delta \lambda$) \\
\hline	
		    & SDSS        & 10-21-2001 & 52203 & 3106  & 3800-9200 & 1800  \\
	J013656.31-004623.8	    & SDSS        & 09-05-2010 & 55444 & 6305 & 3650-10300 & 1800 \\	
  (J0136)      & SDSS & 12-21-2014 & 57012 & 4500 & 3650-10300 & 1800 \\
	$z=1.7154$     & SDSS & 11-24-2017 & 58081 & 3600 &	3650-10300 & 1800 \\
		    & HET/LRS-2  & 10-29-2021 & 59516 & 1210 & 3700-6840 & 2500 \\
		    & P200/TripleSpec  & 10-16-2021 & 59503 & 3600 & 10000-25000 & 2700 \\
\hline 
		    & SDSS        & 05-14-2007 & 54234 & 3406 & 3800-9200 & 1800 \\ 
J123820.19+175039.1 & SDSS  & 04-18-2012 & 56035 & 2702 & 3650-10300 & 1800  \\
	(J1238) & P200/TripleSpec  & 03-05-2018 & 58182 & 900 & 10000-25000 & 2700 \\
$z=0.453$	&  LJT/YFOSC     & 03-21-2018 & 58198 & 900 & 3400-9000 & 380 \\
		    & HET/LRS-2 & 04-02-2021 & 59306 & 480 & 3700-6840 & 2500 \\
\hline 
		    & SDSS        & 04-13-2005 & 53473 & 3700 & 3800-9200 & 1800 \\ 
J125942.79+121312.6 & SDSS  & 02-22-2012 & 55983 & 2702 & 3650-10300 & 1800  \\
	(J1259)	  & HET/LRS-2 & 04-08-2019 & 58216  & 900 & 3700-10300 & 2500    \\
	$z=0.7517$	& LJT/YFOSC        & 04-28-2018 & 58601 & 3000 & 3400-9000 & 380    \\
		     &  LBT/MODS     & 02-15-2020 & 58894 & 1800 & 3300-10300 & 2200 \\
		   & HET/LRS-2 & 04-04-2021 & 59308 & 1707 & 6500-10300 & 2500 \\
\hline
                    & SDSS        & 05-13-2005 & 53503 & 2700 & 3800-9200 & 1800  \\
	J134444.32+315007.6	 & SDSS      & 03-12-2013 & 56363 & 2700 & 3650-10300 & 1800  \\
	(J1344) & LJT/YFOSC        & 04-16-2018 & 58224 & 1500 & 3400-9000 & 380 \\
             $z=1.44$      & HET/LRS-2 & 01-06-2019 & 58489 & 900 &	3700-6840 & 2500 \\
		    & HET/LRS-2 & 06-19-2021 & 59384 & 1710 &	3700-6840 & 2500 \\
\hline
\end{tabular}
 \label{table1}
\end{table*}

\section{Observations and data reduction}

\subsection{Candidate selection}

We began with the selection of BAL-acceleration candidates from the LoBAL-variability sample in \citet{Yi19a}, where  a few quasars were found to show apparent velocity shifts within the sampling epochs. Following \citet{Grier16}, we first searched for  quasar candidates of BAL acceleration  from \citet{Yi19a} by requiring monolithic velocity shifts in BAL profile over at least three spectroscopic  epochs for each quasar. However, none of the quasars from that work satisfy the requirement, probably because (1) the  majority of quasars from \citet{Yi19a} have been  observed only two times, (2) the occurrence of BAL acceleration is intrinsically rare, (3)  BAL-profile changes are ubiquitous as reported in previous studies, and/or (4) it is  impossible to assess BAL acceleration from  BAL-transient  events  without sufficient sampling epochs before its disappearance.

Unfortunately, we did not find any quasars with three or more different-epoch spectra from \citet{Yi19a} that displayed monolithic BAL shifts, i.e., they often exhibit large BAL-profile changes and pose great challenges to ascertain the origin of their BAL variability. Therefore, we then searched for  monolithic velocity shifts among these quasars having only two different-epoch spectra and found three candidates, namely SDSS J134444.32+315007.6,  SDSS J123820.19+175039.1, SDSS J125942.79+121312.6 (hereafter J1344, J1238, J1259) meeting  the requirement. We also add the strongest BAL-acceleration candidate, namely SDSS J013656.31$-$004623.8 (hereafter  J0136), from \citet{Grier16} into our final BAL-acceleration list, due partly to the presence of LoBAL species which was mentioned only in passing in that work.

The  early-epoch spectra of the four BAL-acceleration candidates are retrieved directly from the SDSS DR16 archive (see \citealt{Lyke20}); in addition, we also quantify the variability both in the \mgii-BAL  and continuum shape for J0136 following the same prescription in \citet{Yi21}. 
Interestingly, like the  six quasars reported in \citet{Yi21}, J0136 is also undergoing a LoBAL$\rightarrow$HiBAL transformation along with a decrease of dust;  in stark contrast, the other three BAL-acceleration candidates having  small fractional REW changes   become redder in later epochs.  Below we illustrate them in detail with the aid of archived multi-wavelength photometry  and new spectroscopic observations.

\begin{figure*}
\center
  \includegraphics[height=5.6cm,width=18cm,  angle=0]{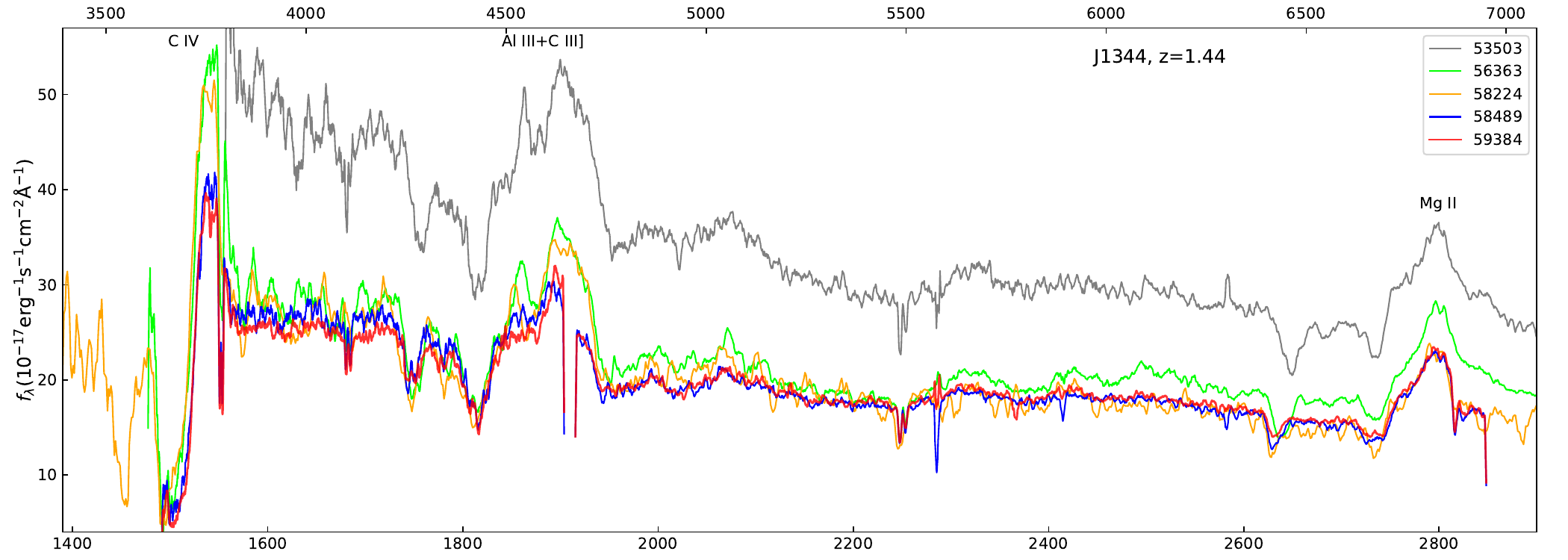}
   \includegraphics[height=5.6cm,width=18cm,  angle=0]{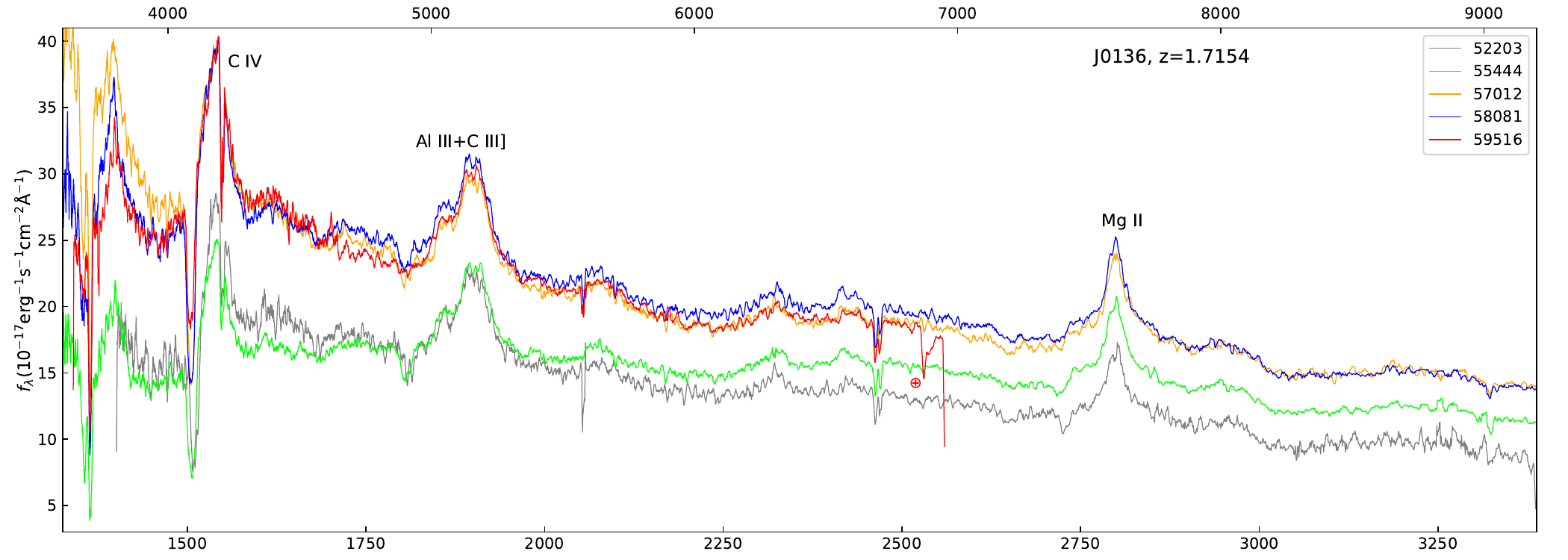}
    \includegraphics[height=5.6cm,width=18cm,  angle=0]{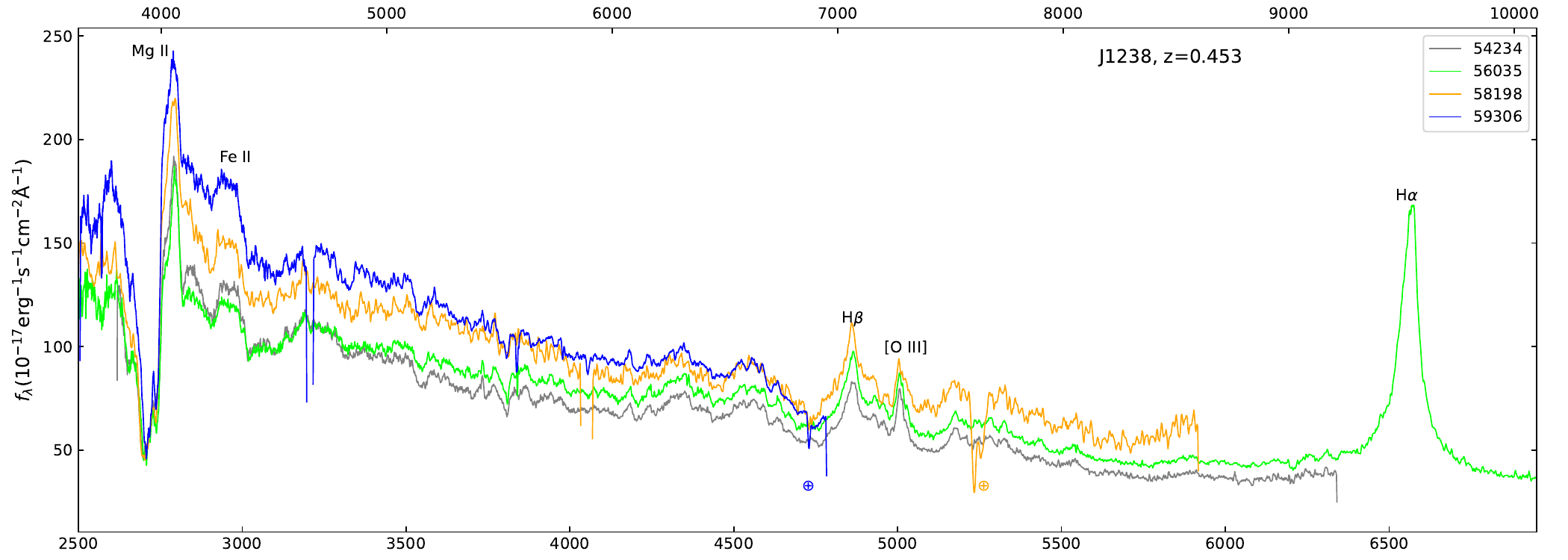}
     \includegraphics[height=5.6cm,width=18cm,  angle=0]{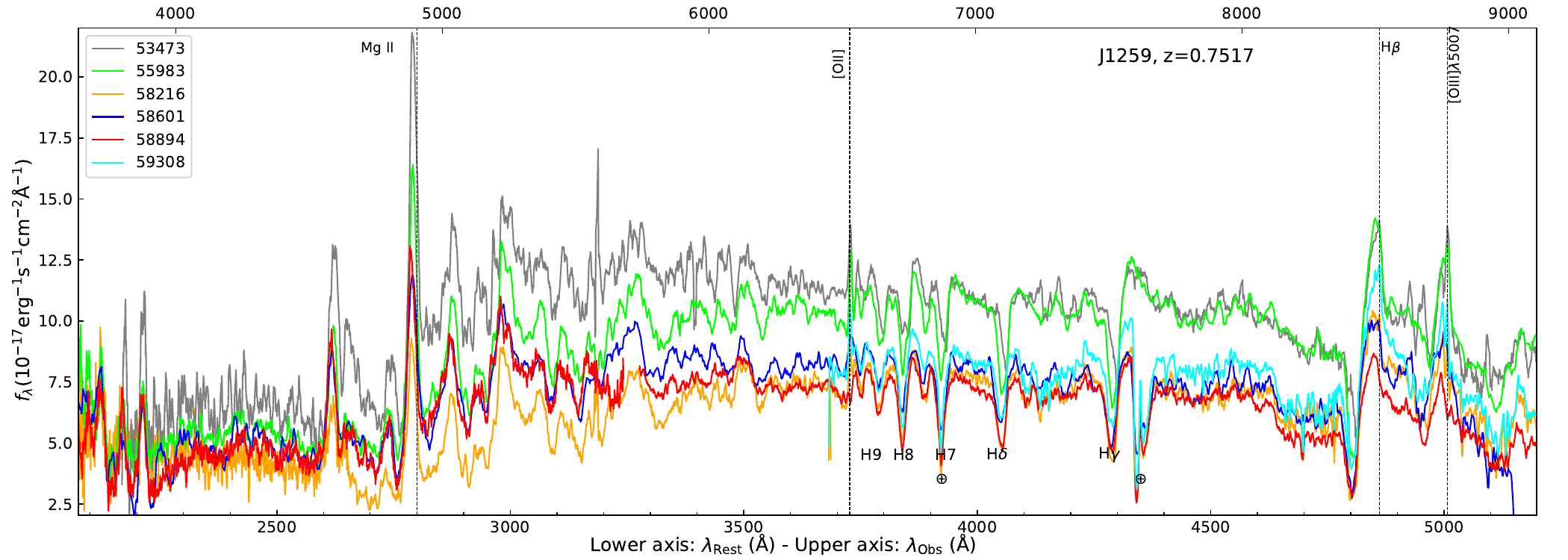}
      \caption{  Multi-epoch optical spectra (smoothed by a 7-pixel  Boxcar filter corresponding to $\sim420 \kms$) of the four quasars having signatures of BAL acceleration, in which the spectral fluxes of J1238 and J1259 are  re-calibrated by $V$-band photometry using the first SDSS epoch as a benchmark. The MJDs of individual observations are given in the upper corner  of each panel. Plus symbols mark telluric absorption. }  
     \label{J1344_5epoch_spec}
\end{figure*}

\subsection{New  observations} \label{new_obs}
To facilitate the assessment of BAL acceleration, we placed these objects into the target list of our spectroscopic monitoring campaign of subsequent BAL variability for individual quasars of interest.  We obtained additional optical spectra for these quasars using the Low-Resolution Spectrograph-2 (LRS2; \citealt{Chonis14}) mounted on the Hobby-Eberly Telescope (HET; \citealt{Ramsey98}). These  newly obtained spectroscopic data were processed with the LRS2 pipeline\footnote{https://github.com/grzeimann/Panacea}, which incorporates an improved procedure for flux calibration with a typical uncertainty of $\sim$15\% (\citealt{Hill21}).  Long-slit  (2.5\arcsec\ in width) spectra were mainly acquired  by the Yunnan Faint Object Spectrograph and Camera (YFOSC) mounted on the Lijiang 2.4 m Telescope (LJT; \citealp{FanY15,WangC19}).  We also obtained another long-slit (1.0\arcsec\ in width) spectrum for the faintest quasar J1259 using the multi-object double spectrographs (MODS) mounted on the Large Binocular Telescope (LBT), which covers a wide wavelength range of 0.33--1.03~$\mu$m. These long-slit data were processed using standard IRAF routines, including bias subtraction, flat field correction, cosmic ray removal, spectral extraction, wavelength identification, and flux calibration.

Near-IR spectroscopic observations were performed with the TripleSpec spectrograph at the Palomar Hale 200 inch telescope (P200/TripleSpec; \citealt{Wilson04}) for  J0136 and J1238. TripleSpec provides a wide wavelength coverage  (0.95--2.46 $\mu$m) at an average spectral resolution of $\sim2700$, allowing simultaneous observations in the J/H/K bands. A slit width of one arcsecond and the ABBA dither pattern along the slit were chosen to improve the sky subtraction during the observations.  We carefully examined the data quality flags and wavelength calibration in each epoch for each quasar, and did not find significant instrumental artifacts from the spectral regions  of interest.

The log of  observations  are summarized in Table 1 and the multi-epoch optical spectra of each candidate are presented in Fig.~\ref{J1344_5epoch_spec} for an overall view.

\section{Spectral measurements and identifications of BAL acceleration }\label{accel_identification}

Following a similar procedure from \citet{Yi19a}, we first model the local continuum of interest by fitting a reddened power-law function to the relatively line-free spectral regions identified through visual inspection for each quasar. The BALs and BELs are then normalized by the fitted continuum from each epoch  for each quasar. Finally, the quantities such as REW are measured by the continuum normalized spectra over the epochs, which allow us to quantify the BALs and BELs and the related time variability therein for each quasar.

We highlight that  the REW alone may fail to quantify BAL variability, especially in cases where BALs show velocity shifts over different epochs but lack significant changes in their profiles,  leading to an approximately equal REW for the two BALs chosen from two different epochs. 
This issue  must  be addressed before performing a further analysis of BAL acceleration, given that one would naturally expect to see velocity shifts of the BAL over multiple epochs from   BAL-acceleration candidates.  Therefore,  it is mandatory to examine the  BAL-profile variability in detail and, whenever possible, to identify  independent or correlated BAL subflows from epoch to epoch  for each quasar.  Such an investigation is  important  given the diversity of BAL-profile variability that may reflect different-velocity components in a single BAL trough arising from largely different physical regions (e.g., \citealp{Arav15,Yi19a,Yi21}). This is also one of the most challenging issues for the  analysis of BAL variability.

Ideally,  high-resolution and high-S/N spectroscopy can provide powerful diagnostics to probe the detailed variability for a given BAL trough, but in reality it is almost impossible to obtain multi-epoch, high-resolution spectra for the  majority of quasars given their faint  apparent magnitudes.  We circumvent this challenge by taking  advantage of  well-separated, intermediate-resolution spectra for each quasar selected. All but the LJT/YFOSC spectra  displayed  below are smoothed by a 3-pixel Boxcar filter ($\sim 200 \kms$) for optimal visual inspection.

\subsection{The criteria of BAL acceleration }

\begin{figure}
\centering
  \includegraphics[height=7cm,width=8.7cm,  angle=0]{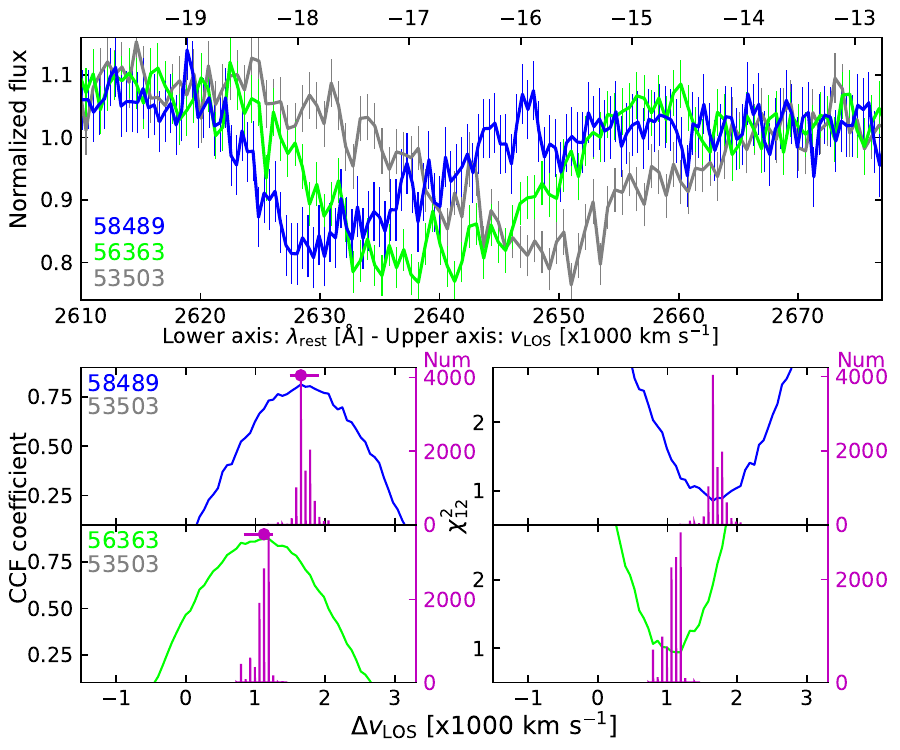}
      \caption{   Demonstration of  monolithic velocity shifts for the \mgii\ BAL in J1344  over three  well-separated  epochs (top) via the CCF (bottom left) and $\chi_{12}^2$ (bottom right) analyses, in which the bar histograms correspond  to the peaks generated from 10000  Monte Carlo simulations, with the dot and error bars depicting the median value of the histogram and  the 90\%  percentile confidence level. Both analyses yield a closely similar velocity shift in each time interval. }  
      \label{J1344MgII_Chi2_vs_CCF}
\end{figure}

As mentioned in the Introduction, identifications of  BAL-acceleration events from apparent velocity shifts are often challenging due to the complex nature of BAL winds.   This issue can be alleviated  to some extent by  placing   stringent criteria to select BAL-acceleration candidates (\citealt{Grier16}); however,  the requirement of a monolithic shift for BAL acceleration in their work, such that the entire BAL trough is clearly shifted but remains generally  unchanged in profile, could potentially miss  cases of genuine BAL acceleration,  given the prevalence of BAL-profile variability both on long and short timescales (e.g., \citealp{Capellupo11,Yi19a,Helmer19}).  To further mitigate this issue and search for potentially more BAL-acceleration cases, we require three  prerequisites:   (1) at least three different  epochs spanning more than one rest-frame year are required for a quasar, (2) the  BAL of interest is similar in profile  and has an overlap in velocity from epoch to epoch, and (3) the BAL  of interest must be free of strong  overlapping absorption  produced by different ions.   Based upon the above preconditions, we  classify  tentative, strong, and compelling cases of BAL acceleration using the following  criteria.

\begin{enumerate}      
 \item 
Tentative  case:  the entire BAL trough of interest is characterized by a monolithic velocity shift, or the same BAL subflow  is persistent  and shows a velocity shift  over  multiple  epochs.  

\item 
 Strong case:  the  BAL velocity  exhibits a monotonic change   over at least  three  consecutive  spectroscopic  epochs.  

\item 
Compelling case:  in addition to  criterion-2,  the measured velocity shifts fall into the same linear relation with time interval. 
\end{enumerate}
      
Following the  prescription of \citet{Grier16}, we perform a cross-correlation function (CCF) analysis to measure the monolithic velocity shift between two different epochs for a BAL of interest. Specifically, we run the cross-correlation analysis  for each of the 10000 iterations via a Monte Carlo approach (see the bottom-left panel of Fig.~\ref{J1344MgII_Chi2_vs_CCF}).

Then, we adopt the median of these  cross-correlation peaks as the best velocity shift, with an error bar  depicting the  90\% percentile confidence level.  We also performed a similar analysis (Monte Carlo simulations) for the BAL of interest using the reduced $\chi_{12}^2$ (see \citealt{Yi19a} for its definition). This exercise  found that the velocity shift and uncertainty are in excellent agreement with those derived by the CCF analysis. For simplicity, we adopt  the CCF analysis as a standard routine to  measure the BAL-velocity shifts and to quantify their uncertainties. We identified one compelling, one strong, and two tentative cases of BAL acceleration  based on the above criteria.

\subsection{ J1344: a case with  compelling evidence for BAL acceleration}

\begin{figure*}
\center
  \includegraphics[height=9.5cm,width=12cm,  angle=0]{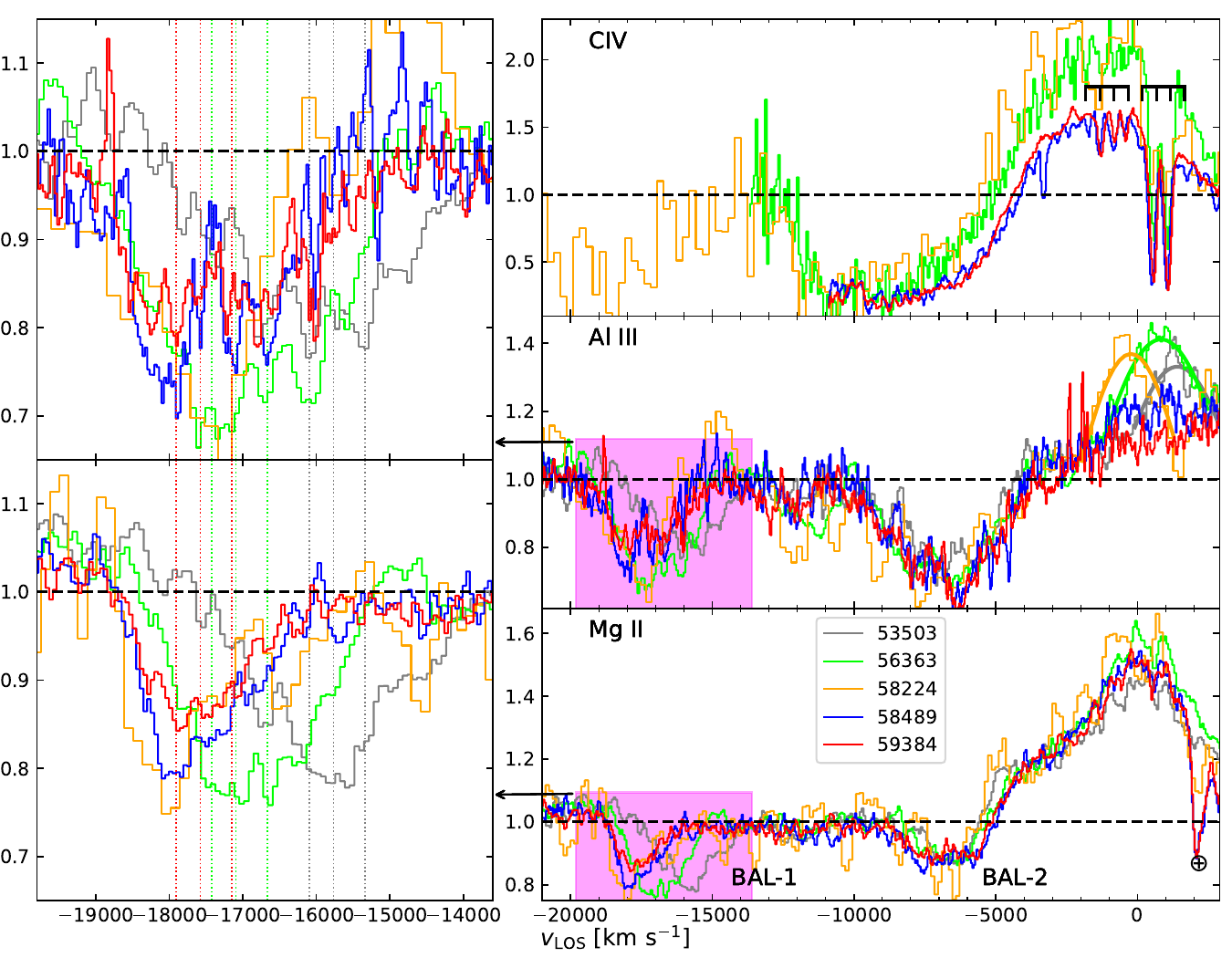}
    \includegraphics[height=9.5cm,width=5cm,  angle=0]{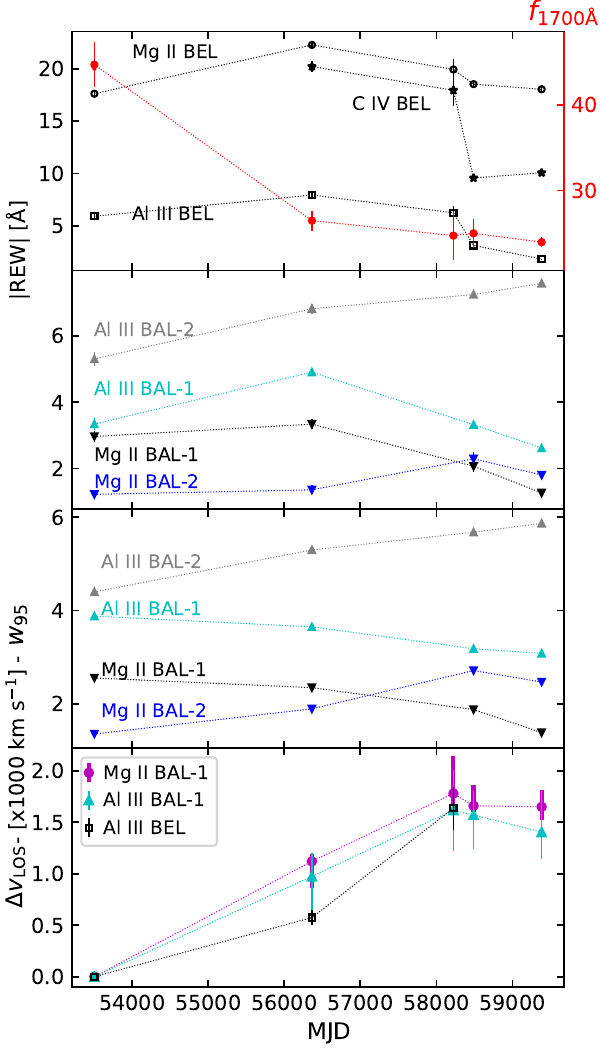}
      \caption{  Left: Zoom-in profiles of the BAL-1 in  \aliii\ and \mgii, in which a set of three blueward components of the \aliii\ and \mgii\ doublets  (gray/green/red vertical dotted lines) with two fixed velocity separations is identified in each of the three spectra with relatively high S/N,  indicative of the same BAL-1 subflow persisting over the epochs.  Middle: Continuum normalized spectra of the group-1 quasar J1344, in which two distinct BALs are labeled with BAL-1 and BAL-2.  The comb-like symbols in the  middle subpanel indicate  radiative-acceleration, line-locking signatures of \civ. The Earth symbol in the bottom-left subpanel indicates telluric absorption. The three  bell-like  profiles depict the one-Gaussian fits to the  apparent  \aliii\ BEL peaks.   Right: MJD vs. velocity shifts (relative to the first epoch), trough widths ($w_{95}$; trough width at  the 95\% continuum level),  REWs  (the \aliii\ emission peak  nearly disappeared  at MJD=59384), and continuum flux at $\lambda_{\rm rest}=1700$~\AA, respectively.    }  
      \label{J1344MgII_aliii_normspec} 
\end{figure*}

J1344 is one of the quasars having two distinct BALs from the sample of \citet{Yi19a}; moreover,  one of the BALs contains a  variable region  based on the measurements from the two different-epoch SDSS spectra as recorded in the catalog. A subsequent  analysis of this quasar using the two SDSS spectra was performed by \citet{LuW20}, where they  found large velocity shifts traced by both \aliii\ and \mgii\  for the high-velocity BAL, and hence  speculated that it could be associated with  actual BAL acceleration. However, as mentioned above, at least three spectroscopic epochs are required to assess the possibility of BAL acceleration in a robust manner.  To achieve this goal, we have obtained  three additional spectra by HET/LRS-2 and LJT/YFOSC.  Although  the YFOSC spectrum  at MJD=58224 has a low spectral resolution and signal-to-noise ratio (S/N),  a reliable velocity shift can be measured  via the CCF approach; moreover,  it is  helpful for visual inspection and can be used for  flux calibration of the HET/LRS-2 spectrum obtained at MJD=58489,  when assuming that  there is  negligible variability in continuum over such a  short time interval.  In combination with  the   well  time-separated  optical spectra  spanning  nearly two decades, we can now  investigate BAL acceleration in a robust manner for this quasar. One can  see from Fig.~\ref{J1344MgII_aliii_normspec} that there are some interesting  features from the normalized spectra: (1) The high-velocity BAL-1 and low-velocity BAL-2 are detected  in \mgii, \aliii, and \civ\  over  the sampling epochs; (2) The  BAL-1 profiles  are similar over the epochs either in \aliii\ or \mgii\ except for small changes in trough depth; (3)  The BAL-1 velocity  increased monotonically from MJD=53503 to 58224 and then appears to level off or perhaps  decelerate  after that epoch; (4) The \mgii\ BAL-2 remains unchanged in  trough depth  over the epochs, despite a large increase in  trough width and a slight increase in centroid velocity;  (5) A third \aliii\ BAL with $v_{\rm LOS}\sim-11000 \kms$ is  significantly detected only at MJD=56363, for which we  cannot assess its  variability in detail. 

\begin{figure*}
\center
  \includegraphics[height=9cm,width=12cm,  angle=0]{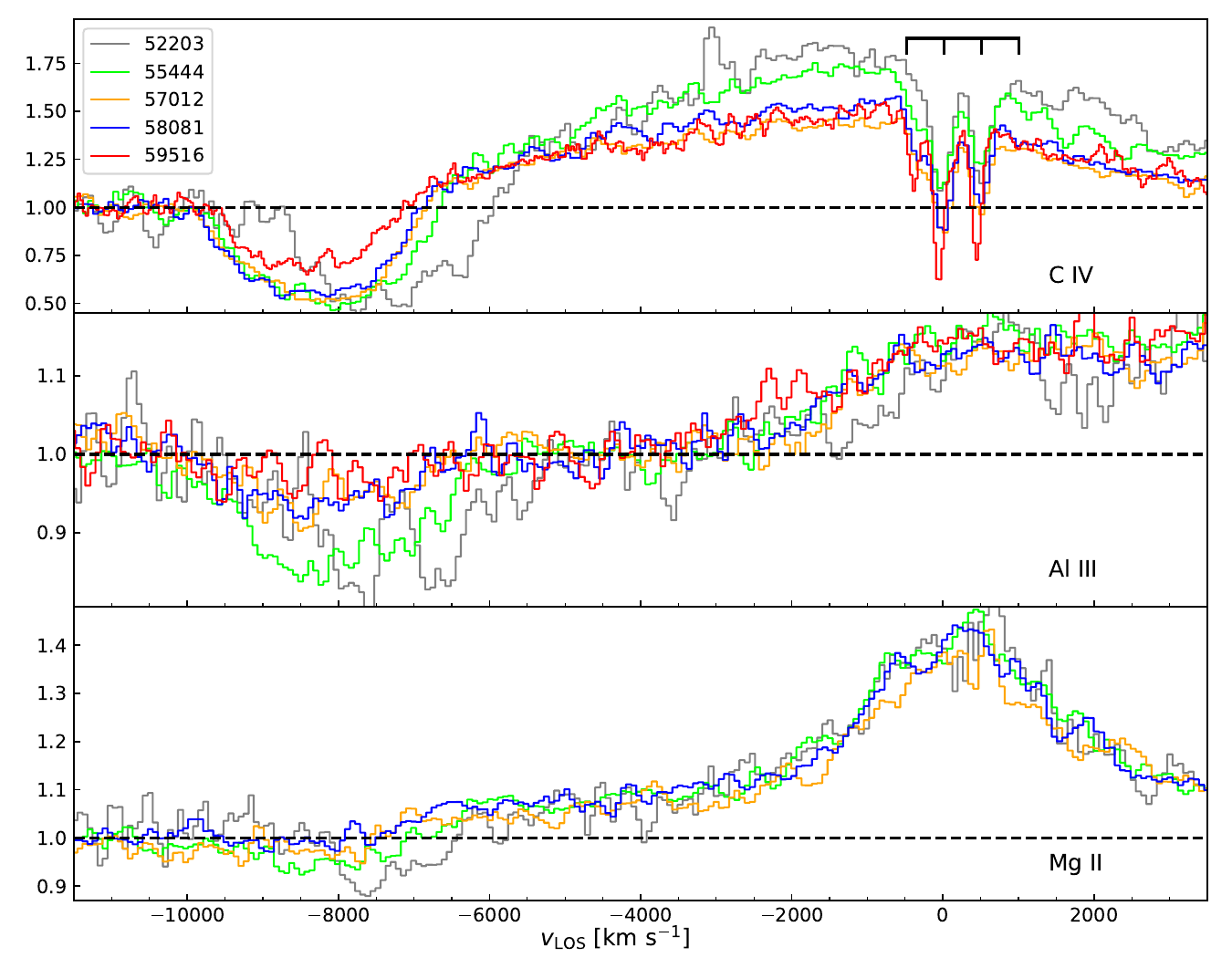}
  \includegraphics[height=9cm,width=5cm,  angle=0]{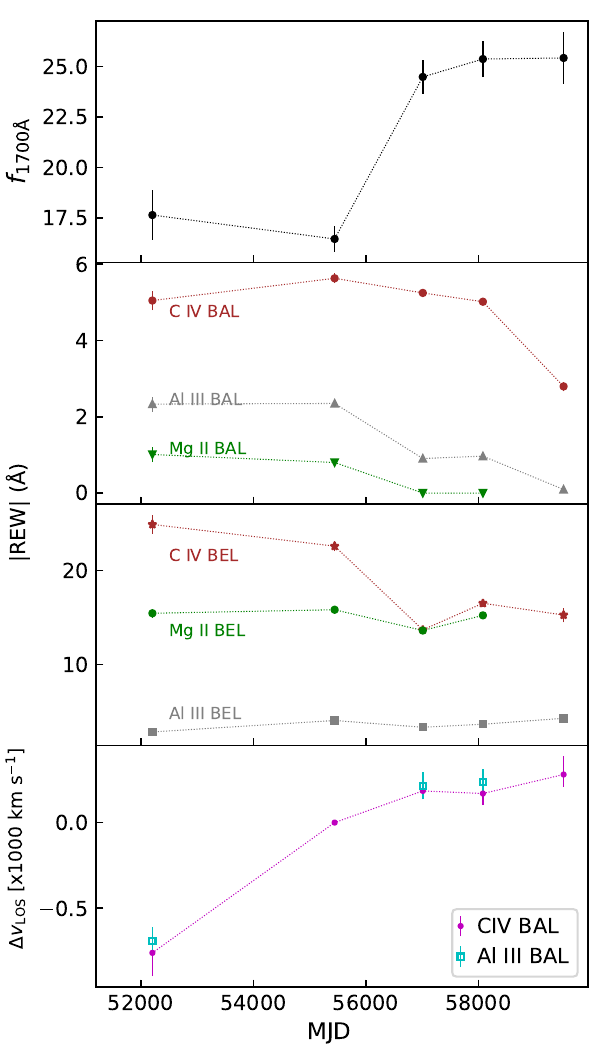}
      \caption{   Left: Continuum-normalized spectra of the group-1 quasar J0136 over five epochs. The   comb-like symbol, again, indicates a line-locking feature in \civ.  Right: MJD vs. velocity shifts of the \civ\ BAL relative to MJD=55444,  REWs for the BALs and BELs, and continuum flux at $\lambda_{\rm rest}=1700$~\AA, respectively, which depict the time-variability  behaviors of these quantities.   The \civ\ BAL profiles are  identical (CCF coefficient $>$0.9) to that from  MJD=55444   and the velocity shifts  in each interval measured by the two different methods are consistent within measurement  errors. The \mgii\  was caught on the disappearance as opposed to the persistence of \civ\ from the BAL flow.   } 
      \label{J013656AliiiNormFlux}
\end{figure*}

Using the first epoch spectrum as a benchmark, we separate the five epochs into four time intervals in an ascending order.  To measure the velocity shifts in a  robust manner, the  \mgii\ BAL-1 that has a higher S/N than the corresponding \aliii\ BAL-1 is treated as the main tracer  for the cross-correlation analysis.  The measured velocity shifts are  $1120_{-260}^{+70}$, $1780_{-60}^{+360}$, $1670_{-65}^{+200}$, $1650_{-130}^{+260}$ $\kms$ over the four intervals via this method.  Similarly, performing a CCF analysis to the  \aliii\ BAL-1  yields  $980_{-430}^{+230}$, $1620_{-390}^{+390}$, $1570_{-330}^{+230}$, $1410_{-260}^{+400}$ $\kms$ for the four time intervals, respectively (see  the MJD vs. $\Delta v_{\rm LOS}$ panel in Fig.~\ref{J1344MgII_aliii_normspec}).   The velocity shifts derived by the two different methods are consistent within the error bars,  reinforcing  the argument for BAL acceleration. In combination with the same linear relation  over the two consecutive  time intervals before MJD=58224, we believe that the observations provide  compelling evidence for BAL acceleration.  On the other hand, by  closely examining   the \aliii\ BAL-1 profiles,  we identified the same subflow with three absorption peaks presumably linked to three different-velocity, blueward components of   \aliii\ doublets from the three  spectra with relatively high S/N (see left subpanels of  Figure~\ref{J1344MgII_aliii_normspec} for details). This identification is further supported by blueward components of   \mgii\ doublets  at the same velocities corresponding to the \aliii\ BAL-1 subflow  in each epoch  (see the vertical dotted lines in  Figure~\ref{J1344MgII_aliii_normspec}), although a shorter-velocity split of the doublet in \mgii\ than in \aliii\ makes it  somewhat uncertain.  
Notably, the velocity shifts traced by the subflow are in excellent agreement with those derived from the entire BAL via the CCF method over the three epochs, reinforcing the argument for acceleration.

For  the analysis of BEL variability,  a velocity range of $-2600<v_{\rm LOS}<2500 \kms$ is chosen  to  characterize  their core-emission features based on visual inspection of the \civ, \aliii, and \mgii\ BELs.  Note that \feii\ emission contributes only slightly to the core portion of the \mgii\ BEL. Surprisingly,   the \aliii\ BEL nearly disappeared by MJD=59384 while the \mgii\ BEL varied only slightly in fractional REW over the epochs. In addition, the \civ\ BEL experienced a sharp drop in REW after MJD=58224. Such dramatic variability of \civ\ BEL has been rarely reported  (\citealt{Ross20}).

Regarding  the evolution of BALs, it is clear that the BAL-1 and BAL-2 REWs  exhibit an overall opposite time-variability pattern after MJD=56363 (see the REW vs. MJD panel in Fig.~\ref{J1344MgII_aliii_normspec}), which may be related to  transverse motions caused by multiple streams moving across our LOS or acceleration/deceleration events  (see Section~\ref{discussion_sec}).   Interestingly,  the \aliii\ BAL-1  and BEL become  weakened  after MJD=56363,  while the \aliii\ BAL-2 and BEL show an opposite  time-variability trend    in REW  after MJD=56363 (see the MJD vs. REW panel of Fig.~\ref{J1344MgII_aliii_normspec}), suggestive of the BEL having a stronger link to BAL-1 in \aliii.  To provide additional  diagnostics for the analysis of the observed BAL/BEL variability, we  added  a panel of MJD vs. $f_{\rm 1700 }$ (the continuum flux at $\lambda_{\rm rest}=1700$~\AA) to Fig.~\ref{J1344MgII_aliii_normspec} for comparison. Clearly, the time-variability pattern of $f_{\rm 1700 }$ is opposite to those of  the BALs/BELs from MJD=53503 to 56363, such that  the BALs/BELs strengthened when the continuum $f_{\rm 1700 }$ dimmed by a factor of $\sim$2; in addition,  the BAL-1 and BELs (in  \aliii\ and \mgii)  show a  similar time-variability trend  to that of $f_{\rm 1700 }$ after MJD=56363.  Conversely, $f_{\rm 1700\AA}$ and velocity shift ($\Delta v_{\rm LOS}$) exhibit an opposite time-variability pattern before MJD=58224; in addition, the BAL-1 and BAL-2 show an opposite time-variability pattern  in trough width at the 95\% continuum level ($w_{95}$) over the five epochs.  Implications of these observational results  will be discussed in Section~\ref{discussion_sec}.

\subsection{J0136: a case with  strong evidence for BAL acceleration}

 J0136 was considered the strongest candidate of acceleration in  \citet{Grier16}  based on the analysis of the \civ\ BAL profiles over different epochs.  But the \aliii\ BAL was mentioned only in passing in their work due to  its  shallow depth. Through a careful examination of the different-ion absorption features in velocity space, we found  that this BAL absorber consists of \siiv, \civ,  \aliii, and  \mgii.  However, we did not explore \siiv\ in this work as it lies at the CCD blue edge. The \mgii\ absorption was detected  only in the first two epochs and completely disappeared in later epochs. Such a phenomenon, in which lower-ionization BAL species disappear faster than higher-ionization BAL species at the same velocity, has been also reported in other quasars from the literature (see \citealp{WangT15,Yi21}).

Through a detailed comparison between the \civ\ and \aliii\ BALs over five epochs, we identified  a number of interesting properties.  (1) The  centroid velocity of the \aliii\ BAL   increases monotonically with time, such that  the velocity shifts relative to MJD=55,444 are $-690\pm$70, 210$\pm$70, and 230$\pm$70 $\kms$ over the three time intervals (see cyan  squares from the MJD vs. $\Delta v_{\rm LOS}$ panel in Fig.~\ref{J013656AliiiNormFlux}), respectively, consistent with those ($-760_{-153}^{+101}$, $185_{-33}^{+52}$, $170_{-68}^{+68}$) derived from the CCF analysis applied to the entire \civ\ BAL trough.  (2) The \aliii/\mgii\ BAL components  completely disappeared while the \civ\ BAL component decreased by a factor of 2 in REW from the first to last epoch. (3) The \civ\ BEL decreased by a factor of $\sim$2  in REW while the \aliii/\mgii\ BELs remained generally unchanged over the epochs. (4) The \civ\ BAL and BEL  have a similar  time-variability pattern characterized by an overall decrease of REW with time. (5) The \aliii\  trough  becomes wider after the first epoch,  consistent with the broadening effect from acceleration (see Section~\ref{discussion_accel_mechanism}),  although we cannot rule out other possibilities.  
Unlike J1344, the \aliii\ BAL in J0136 completely disappeared while its  \aliii\ BEL strength  remained generally unchanged  over the epochs.   Regarding the relation of the continuum and BEL variability, one has to consider the possibility of a time delay between them. However, a full investigation of the BEL variability requires higher-cadence spectroscopic  observations on longer timescales, which is beyond the scope of this work.

J0136  shows kinematic acceleration signatures  traced by both the \civ\ and \aliii\ BAL  troughs,  in which the latter completely disappeared by MJD=59516 characteristic of a LoBAL$\rightarrow$HiBAL transformation in tandem with the acceleration.  Examining  the MJD vs. $f_{1700}$ panel, the continuum flux density at $\lambda_{\rm rest}=1700$~\AA\  brightened by a factor of $\sim$2 during the BAL transformation, a variability pattern that was also reported  in another LoBAL quasar undergoing the same  transformation (\citealp{Yi22}). However, the variability relation between velocity shift and continuum flux    is unclear for J0136, given  both the same and opposite   trends observed over different time intervals, such that the largest velocity shift occurred in the interval with no or small-amplitude continuum variability, while the largest-amplitude continuum variability was detected in the interval having only a small velocity shift along with the \mgii\ BAL disappearance. These results will be discussed in Section~\ref{discussion_sec}. 

\begin{figure*}
\center
  \includegraphics[height=9cm,width=12cm,  angle=0]{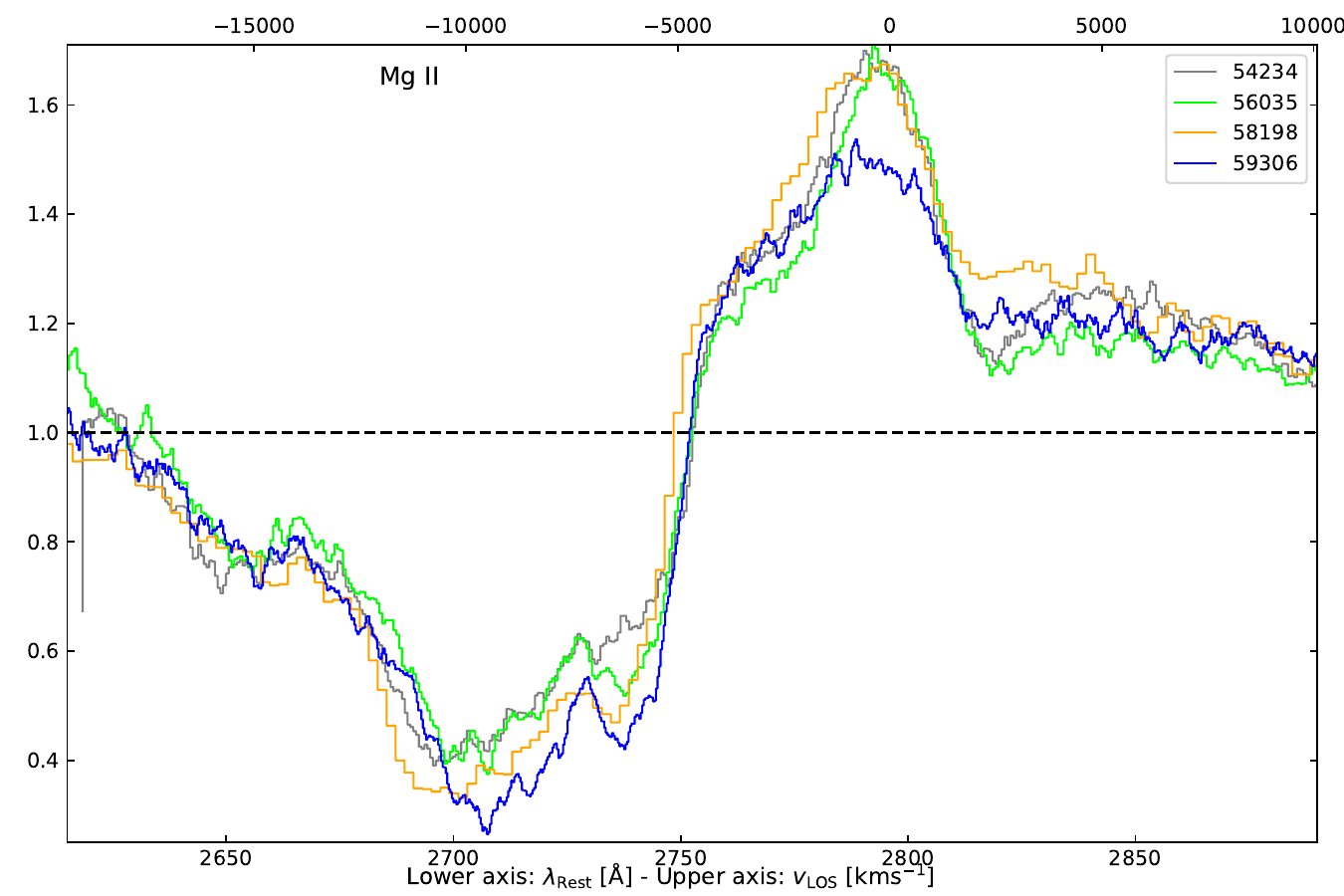}
  \includegraphics[height=9cm,width=5cm,  angle=0]{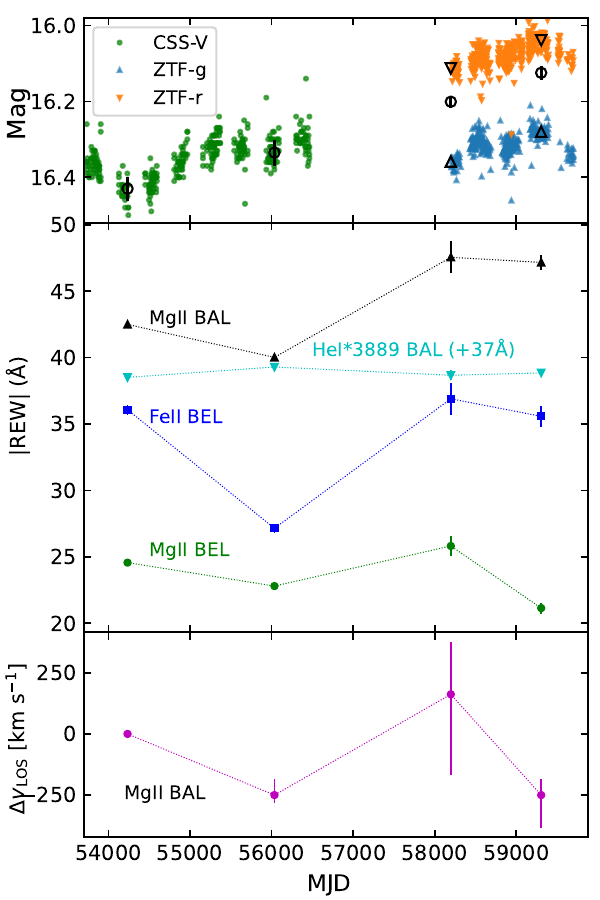}
      \caption{   Left: Continuum normalized spectra of the group-2 quasar J1238 for the \mgii\ BAL/BEL over four epochs.  Right: MJD vs. velocity shift of the \mgii\ BAL,  REW for the BALs and BELs, and photometric magnitudes, respectively, which depict the time-variability patterns of these quantities. Unlike the \mgii\ BAL which has at least three subflows, the \hei\ BAL is detected only at $v_{\rm LOS}\sim6000 \kms$ (see Section~\ref{near_IR_spectroscopy}) and offset above by 37~\AA\  for clarity. The positive/negative velocity shifts detected over the  epochs may be linked to BAL acceleration/deceleration events for this quasar.   }
  \label{J1238spec3_NormFlux}
\end{figure*}

\begin{figure*}
\center
  \includegraphics[height=9cm,width=12cm,  angle=0]{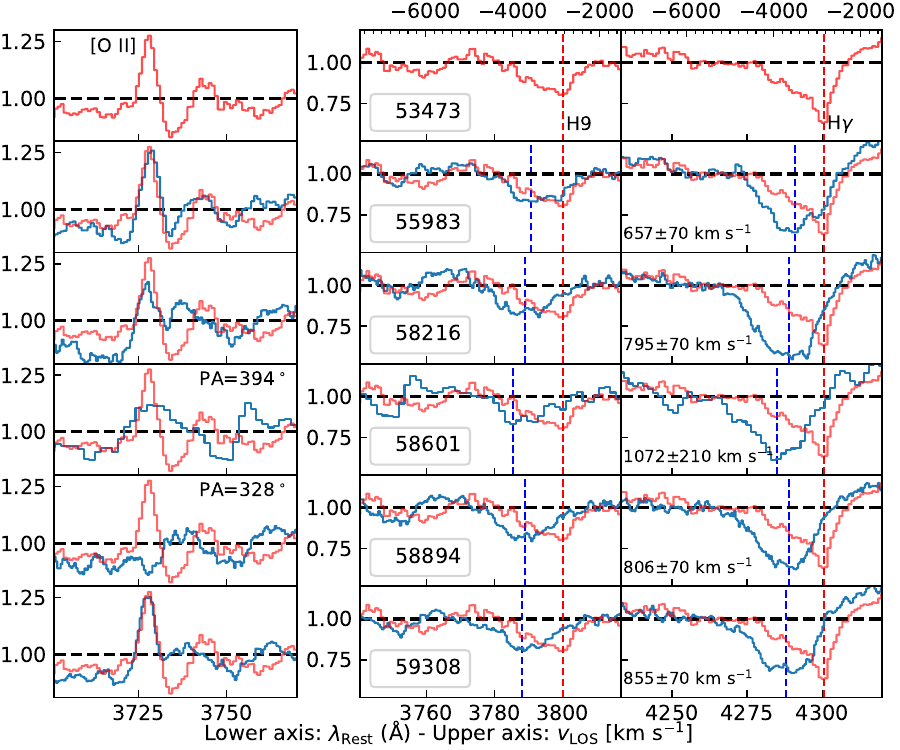} 
  \includegraphics[height=9cm,width=5cm,  angle=0]{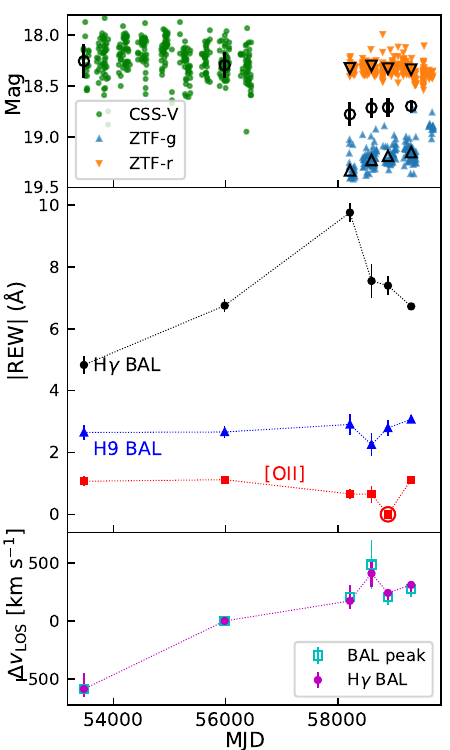}
      \caption{  Left: Continuum normalized spectra of the group-2 quasar J1259 over six epochs. The vertical red/blue dashed lines refer to the peak absorption at the first/other epochs, from which one can directly see how the velocity shift varies with time using the first epoch as a benchmark (the first spectrum is superimposed on all successive epochs and the positions of the red and blue dashed lines in H9 are fixed by those of \hg). Apparently, the H9  remains generally unchanged in REW while the $\hg$ BAL appears to vary from epoch to epoch.  Right: MJD vs. velocity shift of the \hg\ BAL relative to MJD=55983,  REWs of different ions, and photometric magnitudes, respectively. The apparent  disappearance in \oii\ emission at MJD=58894 (highlighted by the red circle) is due to  a rotation (marked by position angle (PA) from the left panels) between the two long-slit spectroscopic observations targeting the quasar core with  a one-sided, off-nuclear \oii\ emission nebula (Yi et al. in prep). }  
      \label{J1259spec6_h9_hg_hb}
\end{figure*}

\subsection{J1238: a case with  tentative evidence for BAL acceleration} \label{J1238_results}

The spectra of J1238 show  that the \mgii\ BAL  profiles are closely similar to each other and contain at least three BAL subflows in each epoch (see the left panel of Fig.~\ref{J1238spec3_NormFlux}). Although each subflow may be composed of many \mgii\ doublets,  we are unable to identify them due to saturation, self-blending, insufficient spectral resolution, and potentially overlapping  with \feii.   The deepest two BAL subflows, at $\lambda_{\rm rest}\sim2700$ and $\sim2735$  \AA, vary in depth and velocity from epoch to epoch, while the shallowest  BAL subflow at $\lambda_{\rm rest}\sim2650$~\AA\ appears not to change in depth.  In addition, the   \hei\ $\lambda$10830  BAL  profile  is closely similar to the \mgii\ BAL trough except for trough depth (see Section \ref{near_IR_spectroscopy}).

The continuum flux density, however, increases monotonically with time over the four epochs as observed by the $V$-band magnitudes (see the black circles in the MJD vs. Mag panel of Fig.\ref{J1238spec3_NormFlux}, in which the last two $V$-band magnitudes are converted from their  ZTF-$g/r$ magnitudes; see \citealt{Jester2005}).  Interestingly, the \mgii/\hei\ $\lambda$3889 BAL strength decreased/increased between MJD=54243 and 56035 while both  increased between MJD=58198 and 59306,  despite the increase of continuum flux over both intervals. Such complex BAL-variability behaviors  are consistent with   previous studies that disfavor the pure ionization-change scenario (\citealp{McGraw17,Yi19a}).

In the analysis of BEL variability, one must keep in mind that the complex emission at $2750<\lambda_{\rm rest}<3000$~\AA\  appears to be dominated by \feii\ rather than \mgii,  given  the generally unchanged \mgii\ BEL  as opposed to  large variability in the \feii\ BEL at $2815<\lambda_{\rm rest}<2900$ \AA\ from MJD=54234 to 58198. Nevertheless, the \mgii\ BEL profile is asymmetric and exhibits a blueshift relative to systemic redshift; in addition, the \mgii\ BEL weakened dramatically from the first to last epoch, perhaps leading to the apparent increase of the \mgii\ BEL blueshift.  The \mgii\ BAL and BEL show an overall opposite time-variability pattern, which will be discussed in Section~\ref{discussion_sec} in conjunction with J0136 and J1344.

Compared to the BAL-1 in J1344, which exhibits a large kinematic shift over the rest-frame 6.6 years, the \mgii\ BAL  in J1238 has much smaller kinematic shifts ($\Delta v_{\rm LOS}$ = $-257_{-33}^{+60}$, $+150_{-330}^{+220}$, $-210_{-130}^{+67}$ $\kms$) relative to the first epoch over the three time intervals after performing the CCF analysis as demonstrated above, perhaps  indicative of  milder acceleration/deceleration events. The evidence for BAL acceleration remains  tentative due to the low spectral resolution at MJD=58198. Alternatively, using the deepest BAL subflow, we  found that its characteristic  velocity has also experienced a decrease$\rightarrow$increase$\rightarrow$decrease process.  These results support  both acceleration and deceleration events occurring at least for the deepest BAL subflow, if not for the entire BAL trough. No clear relations, again,  have been found between  velocity shift ($\Delta v_{\rm LOS}$) and continuum flux, in agreement with J0136. This variability behavior  will be discussed in Section~\ref{discussion_sec}.

\subsection{J1259: a case with  tentative evidence for BAL acceleration} \label{J1259_results}

As reported in the discovery paper from \citet{Hall07b}, the optical spectrum of J1259 is notably characterized by   Balmer absorption features, a rare phenomenon observed in BAL quasars.  Since a BAL seen in a singlet like H9 is naturally expected to be narrower than that from a doublet like \civ,  we treat these Balmer absorption features in J1259 as ``BALs''  throughout this work. Note that these Balmer troughs have an     overlap  to the  \mgii\ BAL in LOS velocity and at least the $\hg$ trough at MJD=58216 is consistent with a bona fide BAL.  
From subsequent  studies of this quasar (e.g., \citealp{ShiX16,Yi19a}),  significant variations were detected for both the Balmer and \mgii\ BALs based on the analysis of the two SDSS spectra (MJD=53473 and 55983). Although  \citet{ShiX16} proposed that BAL acceleration likely occurred in this quasar,  evidence for acceleration remains lacking due to limited  sampling epochs in their work.  With the aid of additional four optical spectra obtained by HET/LRS-2, LJT/YFOSC, and LBT/MODS, we are now able to assess BAL acceleration in considerable  detail by analyzing the similarities and differences among the six different-epoch (over $\sim$9 rest-frame yr) spectra for this quasar.

To alleviate contamination from telluric absorption and blending with other ions (for details see \citealp{Hall07b, ShiX16})  during the analysis, we choose to examine the H9 and $\hg$  absorption  features  since  they consist primarily of a single ion with relatively high spectral  S/N. It is obvious from Fig.~\ref{J1259spec6_h9_hg_hb} that the two BAL  profiles varied most dramatically during the time interval from the first to second epoch, but changed  only slightly over the later epochs.  Therefore we adopt the second-epoch spectrum as the benchmark  when applying the CCF analysis to this quasar.   The CCF coefficient ($\rho$=0.74) is lowest  for the  epoch pair between MJD=53473 and 55983, consistent with our visual inspection that the $\hg$ BAL profile at the first epoch is somewhat apparently  different from those at other epochs.  In addition, we visually examine the absorption peaks across the two different-ion BAL troughs over the epochs, whose differences may trace the velocity shift when assuming the same BAL subflow persistent in each epoch.

The $\hg$ BAL spectral region has the highest S/N  among these Balmer lines,  so it is used  as a benchmark to search for absorption peaks and velocity shifts over the   epochs  for this quasar. In Fig.~\ref{J1259spec6_h9_hg_hb} the red vertical line indicates  the peak absorption at the first epoch and is shown  in the same column  for  easy comparison, while the peak-absorption positions are indicated by the blue vertical lines in the other five epochs.   The H$\gamma$ absorption peak is consistent with that of the H9  BAL  in each epoch. 
Therefore, using the deepest absorption feature as a characteristic velocity of the BAL absorber, we also found  that  the BAL velocity increases monotonically from MJD=53473 to 58601 and  decreases slightly over the later two epochs, indicative  of  BAL acceleration and perhaps deceleration  for this quasar. Quantitatively, the velocity shifts traced by the peak absorption of $\hg$ over the other five successive  epochs are 657$\pm$70, 795$\pm$70, 1072$\pm$260, 806$\pm$70, 855$\pm$70 $\kms$  relative to the first epoch.  The last four velocity shifts are therefore 138$\pm$70, 415$\pm$260, 149$\pm$70, 198$\pm$70 relative to the second epoch, which are in  agreement with those ($173_{-70}^{+140}$, $410_{-117}^{+117}$, $242_{-69}^{+13}$, $311_{-69}^{+20}$  $\kms$ relative to the trough at MJD=55983) derived from the CCF analyses to all but the first-epoch spectra  having an overall identical \hg\ BAL profile as tested by the maximum CCF coefficient ($\rho>$0.9 vs. $\rho=$0.74). This result reinforces  the argument for BAL acceleration. However,   the above results provide only tentative evidence of BAL  deceleration,  due to a large uncertainty from the low-S/N,   low-resolution spectrum at MJD=58601.

 \begin{table*}
\centering
\caption{  Measurements of the rest equivalent widths, trough widths, trough-velocity shifts, and acceleration magnitudes   }
\begin{tabular}{lcccccc}
  \hline\noalign{\smallskip}
Name & MJD & REW & REW & REW$|w_{95}$ & $\Delta v|a$ & $\Delta v|a$ \\
 & & (\AA) & (\AA)  & (\AA$|\kms$)  & ($\kms|\acms$) & ($\kms|\acms$) \\
  \hline\noalign{\smallskip}
 && \mgii\ BAL-1 & \aliii\ BAL-1 & \mgii\ BAL-2 &  \mgii\ BAL-1 &  \aliii\ BAL-1  \\
                    & 53503$\blacksquare$        & 2.96$\pm$0.17 & 3.34$\pm$0.18  & 1.2$\pm$0.12 $|$ 1350 & - & -  \\
                    \\
		 & 56363      & 3.33$\pm$0.15 & 4.9$\pm$0.13 & 1.34$\pm$0.14 $|$ 1890 & 1120$_{-260}^{+70}|$1.10$_{-0.26}^{+0.08}$ &   980$_{-430}^{+230}|$0.97$_{-0.42}^{+0.22}$  \\
		 \\
	J1344 & 58224        & 3.1$\pm$0.61 & 3.76$\pm$0.60 &  2.07$\pm$0.52 $|$ - & 1780$_{-120}^{+360}|$1.06$_{-0.08}^{+0.22}$ &  1620$_{-390}^{+390}|$0.97$_{-0.23}^{+0.23}$ \\
	\\
                   & 58489 & 2.05$\pm$0.17 & 7.21$\pm$0.27 & 2.29$\pm$0.21 $|$ 2712 &  1670$_{-65}^{+200}|$0.95$_{-0.03}^{+0.10}$ &  1570$_{-330}^{+230}|$0.89$_{-0.18}^{+0.13}$ \\
                   \\
		    & 59384 & 1.27$\pm$0.11 & 7.60$\pm$0.06 & 1.80$\pm$0.13 $|$ 2466 &  1650$_{-130}^{+160}|$0.79$_{-0.06}^{+0.08}$ &  1410$_{-260}^{+400}|$0.68$_{-0.13}^{+0.18}$ \\
 \hline\noalign{\smallskip}
 && \mgii\ BAL  & \aliii\ BAL & \civ\ BAL  &  \civ\ BAL &    \\
                    & 52203        & 1.01$\pm$0.16 & 2.33$\pm$0.20  &  5.05$\pm$0.25 $|$ 2560 & 760$_{-153}^{+101}|$0.74$_{-0.14}^{+0.09}$ & -  \\
                    \\
		 & 55444$\blacksquare$    & 0.84$\pm$0.10 &  2.35$\pm$0.10 &  5.63$\pm$0.13 $|$ 2887 & - &   -  \\
		 \\
	J0136 & 57012        & - & 0.9$\pm$0.09 &  5.25$\pm$0.08 $|$ 2886 & 185$_{-33}^{+52}|$0.37$_{-0.07}^{+0.10}$ &  - \\
	\\
                   & 58081 & - &  0.97$\pm$0.09 &  5.02$\pm$0.09 $|$ 2684 &  170$_{-68}^{+68}|$0.20$_{-0.07}^{+0.07}$ &  - \\
                   \\
		    & 59516 & - & - & 2.79$\pm$0.12 $|$ 2258 &  281$_{-70}^{+105}|$0.22$_{-0.05}^{+0.08}$ &  - \\
 \hline\noalign{\smallskip}
 && \mgii\ BAL & \hei$\lambda$3889  & \hei$\lambda$10830  &  \mgii\ BAL &  \mgii\ BAL   \\
                    & 54234$\blacksquare$        & 42.50$\pm$0.27 & 1.51$\pm$0.15  & - & - & -  \\
                    \\
		 & 56035      & 40.01$\pm$0.22 & 2.29$\pm$0.14 & - & --257$_{-33}^{+60}|$--0.24$_{-0.03}^{+0.06}$ &   -  \\
		 \\
	J1238 & 58198        & 47.55$\pm$1.20 & 1.67$\pm$0.37 & 88.83$\pm$3.07 $|$ 12350  & 150$_{-330}^{+220}|$0.06$_{-0.13}^{+0.10}$ &  - \\
	\\
                   & 59306 & 47.17$\pm$0.58 & 1.85$\pm$0.17 & - &  --210$_{-130}^{+67}|$--0.07$_{-0.04}^{+0.06}$ &  $a_{34}=-0.55_{-0.09}^{+0.07}$ \\
 \hline\noalign{\smallskip}
 && H9 BAL &  & \hg\ BAL &   \hg\ BAL &   \hg\ BAL   \\
                    & 53473        & 2.65$\pm$0.24 & -  &  4.83$\pm$0.29 $|$ 1707  & 590$_{-70}^{+130}|$0.48$_{-0.06}^{+0.12}$ & -  \\
                    \\
		 & 55983$\blacksquare$      & 2.66$\pm$0.20 &  - &  6.75$\pm$0.22 $|$ 1910 &- &   -  \\
		 \\
	        & 58216        & 2.91$\pm$0.33 & - &  9.76$\pm$0.30 $|$ 2347 & 173$_{-70}^{+140}|$0.16$_{-0.06}^{+0.12}$ &  - \\
	\\
        J1259 & 58601 & 2.26$\pm$0.42 & - &  7.55$\pm$0.55 $|$ 2143 &  410$_{-117}^{+117}|$0.32$_{-0.09}^{+0.09}$ &  - \\
                   \\
		    & 58894 & 2.79$\pm$0.25 & -  &  7.40$\pm$0.32 $|$ 2332 & 242$_{-69}^{+13}|$0.17$_{-0.05}^{+0.01}$ &  $a_{45}=-1.17_{-0.13}^{+0.13}$  \\
		    \\
		    & 59308 & 3.07$\pm$0.13 & - &   6.72$\pm$0.12 $|$ 2161 &  311$_{-69}^{+20}|$0.19$_{-0.04}^{+0.01}$ &  - \\
 \noalign{\smallskip}\hline
\end{tabular} \\
\footnotesize{ Note: (1) The  square  represents the benchmark spectrum  used for the CCF analysis for each quasar. (2)   $w_{95}$ is the  trough width at the 95\%  continuum level. (3) $a_{34}$  and $a_{45}$ are the  deceleration rates derived from the 3rd/4th and 4th/5th epoch pairs, respectively. }
 \label{table2}
\end{table*}

It is worth noting that  the \oii\  emission line in J1259, at first glance,  appears  to have experienced a sudden disappearance at MJD=58894 and a reappearance at MJD=59308 (see  Fig.~\ref{J1259spec6_h9_hg_hb}).   However,  both the IFU data  at  MJD=58216 and 59308 reveal a similar one-sided, off-nuclear \oii\ emission nebula with a projected size of $\sim$25 kpc (Yi et al. in prep).    
Therefore,  a long-slit spectroscopic observation that was targeting the quasar core could  make the \oii\ emission photocenter fall out of the slit with PA=328$^\circ$, leading to the apparent \oii\ disappearance from the long-slit, background-subtracted spectrum at MJD=58894.  On the other hand, the  \oiii\  emission exhibits a  blueshifted, broad-wing feature in each epoch,  suggesting  that   some of the BAL winds may  be closely linked  to  the forbidden-line outflows.   A dedicated study,  especially via spatially  resolved 3D spectroscopy, would be  valuable  to gain unique insights  into the origin of the giant \oii\ nebula;  this effort  is beyond the scope of this work. We  briefly discuss the implications of these  results   in Section~\ref{discussion_sec} after combining the other three quasars, particularly in the context of hot dust and weak radio emission.

\subsection{ An overall view  of BAL acceleration     } \label{overall_view}

\begin{figure}
\centering
  \includegraphics[height=5cm,width=8.7cm,  angle=0]{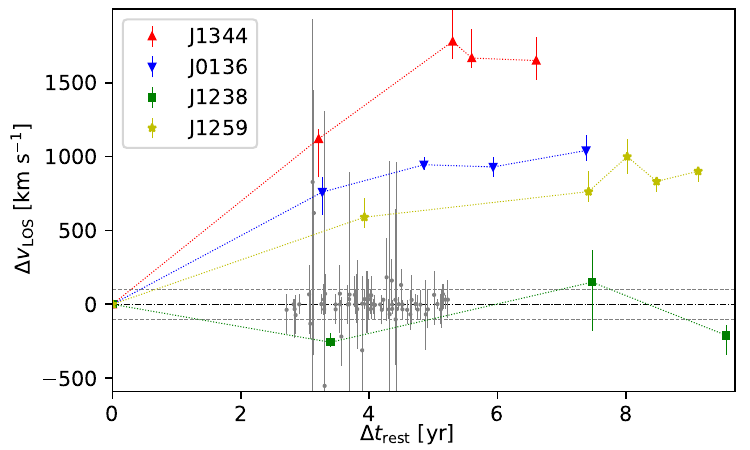}
      \caption{  Comparison between the LoBAL (color) and HiBAL (gray;   from Table 4 of \citealt{Grier16}) samples. The two horizontal dotted lines are the upper limits of  BAL acceleration/deceleration  in \citet{Grier16}  derived by the median values of velocity shifts + 3$\sigma$ positive/negative uncertainties. J1344, J0136, and J1259 all show a similar trend, such that they have a steep rise followed by a plateau-like time evolution  in  BAL-velocity shift.   }  
      \label{cut_velocity_shift_vs_dt_Grier16} 
\end{figure}

The two quasars from group-1 possess multiple BAL-acceleration signatures. J1344 is   the most convincing  example  of BAL acceleration known to date, given that (1) the BAL profiles of interest are   similar  over the epochs except for differences in velocity shift; (2) the velocity shifts over each time interval measured by two different methods are approximately equal; (3) its largest BAL-velocity shift in \mgii\ (1780 $\kms$) is much higher than the median value (98 $\kms$) of positive velocity shifts + 3$\sigma$ uncertainties from Table~4 in \citet{Grier16} (see Fig.~\ref{cut_velocity_shift_vs_dt_Grier16});  and (4) most importantly, our data reveal the velocity shift in J1344 as a linear function of time interval over three well-separated, consecutive epochs, which is highly unlikely caused by BAL-profile variability and other random effects. For J0136 (another quasar  in group-1),  it has all but  property-(4)  listed above, which can be considered a  less robust but still convincing example of BAL acceleration.

In contrast, BAL-acceleration signatures of the two quasars from group-2 are less convincing than those from group-1, due to smaller BAL shifts and the lack of wide-separation  doublets such as \aliii\ that can be well  resolved by the intermediate-resolution spectroscopy. The \mgii\ doublet has a shorter-wavelength separation than \aliii\ and is often unidentifiable from a BAL trough due to saturation and blending with \feii. However, the group-2 quasars still  have  negative BAL-velocity shifts  at a significant level relative to the measurement uncertainties (--260$\pm$70/--270$\pm$70 $\kms$ for J1238/J1259); moreover,  these negative shifts are   almost  3 times higher than the  median value (--102 $\kms$) of negative velocity shifts -- 3$\sigma$ uncertainties  from \citet{Grier16},  supportive of BAL deceleration.  Note that  the smaller the BAL-velocity shift between two epochs, the greater the chance that the shift could be due to velocity-dependent trough depth changes caused by  ionization changes and/or transverse motions, and not bulk acceleration or deceleration.   Table \ref{table2}  provides  an overall view of  the  spectral measurements.

\section{   Multi-wavelength information}

\subsection{The SED properties  } \label{sec_SED}

\begin{figure*}
\center
  \includegraphics[height=5cm,width=8.7cm,  angle=0]{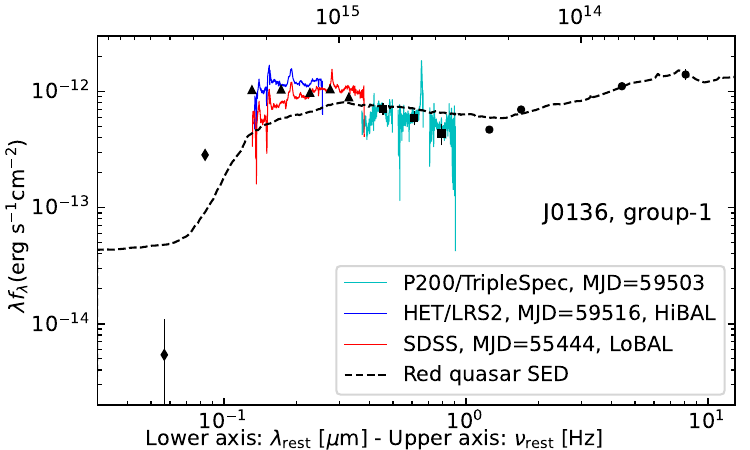}
   \includegraphics[height=5cm,width=8.7cm,  angle=0]{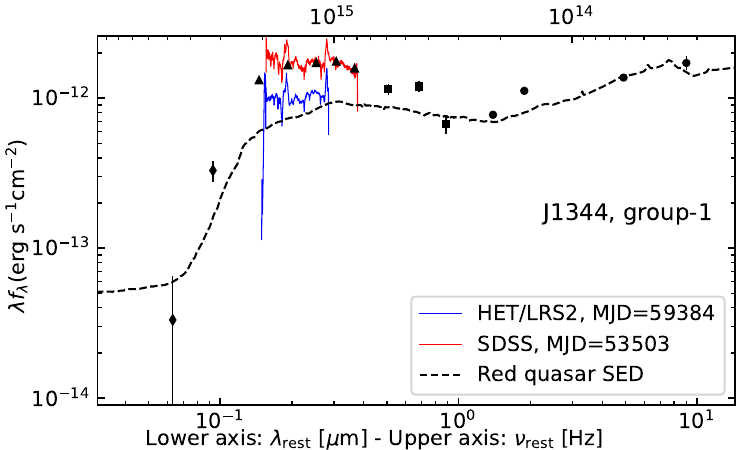}
   \includegraphics[height=4.5cm,width=8.7cm,  angle=0]{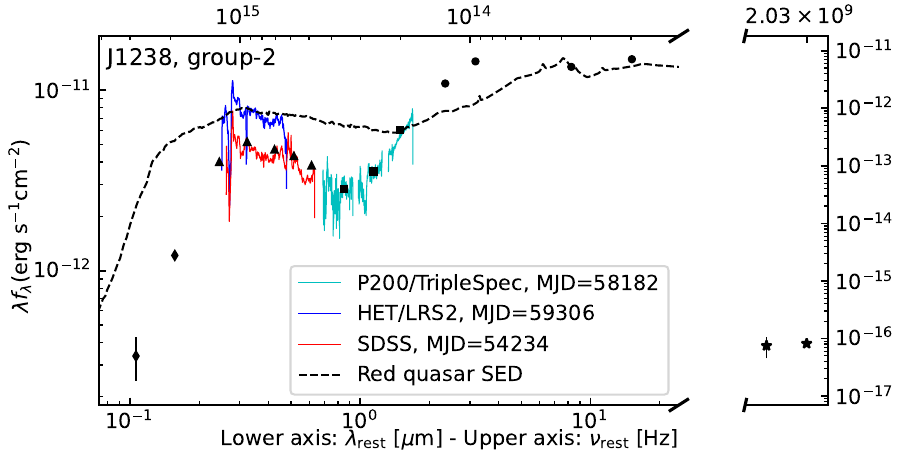}
   \includegraphics[height=4.5cm,width=8.7cm,  angle=0]{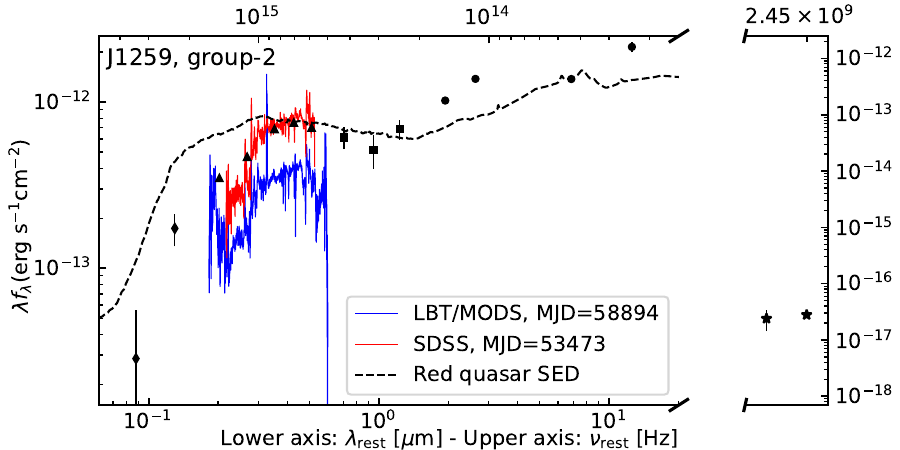}
      \caption{   SEDs of the four quasars, which incorporate data from  GALEX (diamond), SDSS (triangle), 2MASS (square), WISE (circle),  VLASS and FIRST   (star).   The early/late  epoch  spectra are shown in red/blue  for each quasar, complemented with a near-IR (cyan)  spectrum if it exists. J0136 and J1344  (group-1) have BAL-acceleration signatures and are radio quiet;  in addition, J0136 is  undergoing a LoBAL$\rightarrow$HiBAL transformation along with an increase of brightness over the epochs, while J1344 remains to  be in a LoBAL state along with a decrease of brightness.   J1238 and J1259  (group-2) exhibit BAL-acceleration/-deceleration signatures and weak radio emission, with  $f_{\rm 3GHz}/f_{\rm 1.4GHz}=2.5/5.84$ mJy and $0.8/2.0$ mJy for the former and latter, respectively.  The GALEX photometry for J1259 should be treated with caution given its large flag values.  The median SED of red quasars from \citet{Rivera21} is scaled to the W3-band.  }  
      \label{J1344sed_2epoch}
\end{figure*}

This section  explores  the spectral energy distribution (SED) properties for each of the four quasars,  with a particular attention focusing on   variability of the rest-frame UV reddening between two different epochs.  J0136 and J1344 are BAL-acceleration candidates without significant radio detections from FIRST, while J1238 and J1259 exhibit both BAL acceleration and deceleration signatures,  along with weak radio emission and a much redder color as  indicated by the W3/SDSS-$i$ ratio.

Fig.~\ref{J1344sed_2epoch} displays the SEDs of the four quasars, whose photometric data are retrieved from  GALEX, SDSS, 2MASS, WISE, and FIRST sky surveys (see \citealt{Lyke20} and references therein).  To highlight  variability in the rest-frame UV continuum, we display  the early/late epoch spectra in red/blue color  for each quasar.  In addition,  the near-IR spectra obtained by P200/TripleSpec for J0136 and J1238 are  flux-corrected by the photometric data from 2MASS and  included  in their SEDs for comparison.    All the four SEDs have a peak at $\lambda_{\rm rest}\sim 10~ \mu$m;  in particular, the group-2 quasars  have a steeper rise of the SED shape than the group-1 quasars  at $1<\lambda_{\rm rest}<3~\mu$m. 

\begin{figure*}
\center
  \includegraphics[height=8.5cm,width=8.7cm,  angle=0]{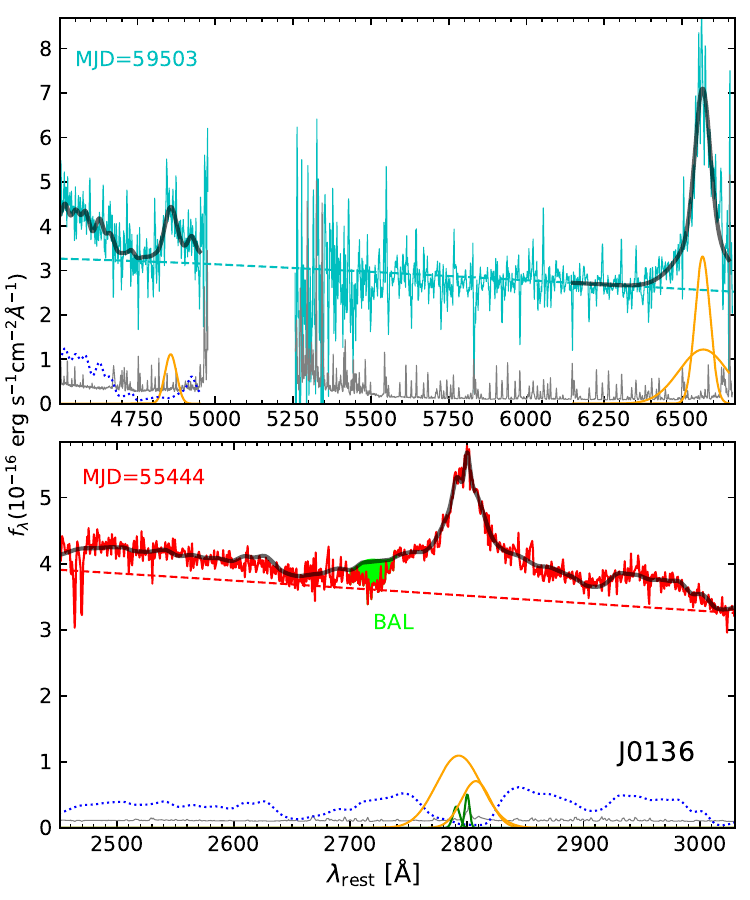}
  \includegraphics[height=8.5cm,width=8.7cm,  angle=0]{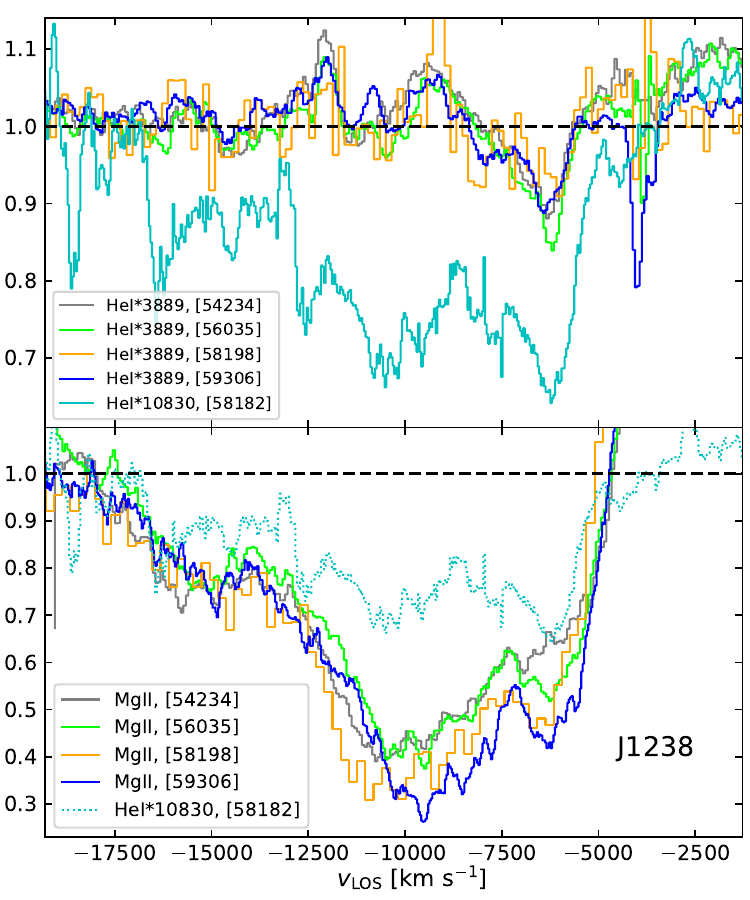}
      \caption{    Left panels: The  spectral fits (thick gray) to the \mgii\ (bottom panel), \hb\ and \ha\ (top panel) emission for J0136, which include the \feii\  (blue dotted), the broad Gaussians (orange), the narrow Gaussians (green), and the continuum fits (dashed). The BAL feature (green) was masked during the fit. The thin grey lines  indicate spectral errors.  The  fits indicate that the  FWHM  (\mgii)  is  $\sim4650 \kms$  in  broad emission as opposed to FWHM (\hb) $\sim2750 \kms$, and that  blueshifted emission/absorption features exist in \mgii\ but not in \hb, suggestive of the former being more affected by outflows.   Right panels: The \hei\ $\lambda10830$ and $\lambda3889$ BAL profiles  for J1238. For comparison, a subpanel of the \mgii\ BAL profile is added  over the four epochs, above which  the  \hei\ $\lambda10830$ BAL is  displayed. Unlike the \hei\ $\lambda10830$ and \mgii\ BALs having  three major subflows, only one \hei\ $\lambda3889$ BAL feature was  detected at $v_{\rm LOS}\sim-6000~\kms$, whose strength remains generally unchanged  over the epochs.  }  
      \label{J123820_HeI_BAL}
\end{figure*}

J0136 is the only one among the four quasars undergoing a LoBAL$\rightarrow$HiBAL transformation and becoming  brighter/bluer  in the rest-frame UV band in later epochs.  Interestingly, such a time-variability trend is also seen in six LoBAL quasars that were caught on  a LoBAL$\rightarrow$HiBAL transformation  (\citealt{Yi21}); in particular, the  \civ, \aliii, and \mgii\  BAL-variability trends in J0136 resemble closely another LoBAL quasar (J0827) that was captured of shedding its dust cocoon, despite the emergence of a new \civ\ BAL in J0827 (see Figure~4 in \citealt{Yi22}). Given the persistence of the BAL in J0136 and the LoBAL$\rightarrow$HiBAL$\rightarrow$non-BAL evolutionary path proposed  in \citet{Yi21},  we argue that the SED of J0136 will become increasingly bluer/brighter, and ultimately like the vast majority of non-BAL blue quasars, peak at the rest-frame UV band as  time passes.  In contrast,  J1344 is still  in a LoBAL state and becomes  conspicuously  dimmer   (a factor of $\sim$2) in the UV continuum  over  the later epochs, although it has an even larger BAL-acceleration magnitude than J0136 (1.14 vs. 0.74 $\acms$; see Table~\ref{table2}).  These results are valuable to  advance our understanding of the BAL nature, which will be  discussed  in Section~\ref{discussion_sec} in conjunction  with the multi-wavelength data at hand.

\subsection{Near-IR spectroscopy for  J1238 and J0136}  \label{near_IR_spectroscopy}

We have performed near-IR spectroscopic observations using the P200/TripleSpec for J1238 and J0136.  The near-IR spectrum of J0136 at MJD=59503  exhibits  strong \feii\ and weak \hb\  emission (see left panels of Fig.~\ref{J123820_HeI_BAL}). Following \citet{Yi22}, we use a model with two Gaussians and one \feii\ component to fit the \ha\ and \hb\ lines.  Our spectral fit reveals that the  \ha/\hb\ emission can be well fitted by two/one Gaussians, when  their FWHMs are tied to each other and the \ha/\hb\ flux ratio is constrained within a range between 2.9 and 3.5 for one set of the  Gaussians during the fit.   The \hb\ FWHM of the broad-emission component is derived to be  $2750\pm120 \kms$; in addition, the \ha\ FWHM measured from a combination of the two broad Gaussians  is $3200\pm50 \kms$, which is still narrower than that measured  from the broad \mgii\ emission ($4650\pm19 \kms$; see Table \ref{table3} for individual components) at MJD=55444. This difference is expected in the context of  stratified structures of the  BLR  for the \mgii\ and \ha\ emission, especially in the scenario where BAL winds are shaping the  \mgii-BEL profile (see an  example from \citealt{Yi22}). Indeed, our spectral fit to the \mgii\ BEL reveals a blueshifted, narrow Gaussian that  well characterizes  the blueshifted-wing emission in \mgii\ (also see  Fig.~\ref{J013656AliiiNormFlux} for a similar blueshifted feature over the other epochs), suggesting the presence of the \mgii\ emission-line outflows and hence making it less robust than \hb\ in the study of BLR physics.  We caution that the  \ha\ and \hb\  FWHMs may be  underestimated due to the CCD edge effect.

\begin{table}
\centering
 \caption{   Spectral fitting results of the BELs  for  J0136 }
 \begin{tabular}{lcccc}
  \hline\noalign{\smallskip}
 & $\lambda_c$  $|$ FWHM &  ...  &  ...  &  ...  \\
 & (\AA) $|$  ($\kms$) & & & \\
  & BG-1  &  BG-2  &  NG-1 &  NG-2 \\
  \hline\noalign{\smallskip}
\mgii\ & 2808 $|$ 3102 & 2793 $|$ 4806 & 2791 $|$ 750 & 2800 $|$ 501 \\
\hb\ & 4860 $|$ 2750 & - & - & - \\
\ha\ & 6564 $|$ 2680 & 6570 $|$ 8346 & - & - \\
  \noalign{\smallskip}\hline
\end{tabular}
\footnotesize{ Note:   BG and NG refer to the broad and narrow Gaussians decomposed from the spectral fitting; $\lambda_c$ for the rest-frame wavelength center of a  specific Gaussian.   }
\label{table3}
\end{table}

The near-IR spectrum of J1238 reveals a dramatic \hei\ $\lambda10830$ absorption feature aligning exactly with the \mgii\ BAL trough in velocity space, whose kinematics is highly similar to the \mgii\ BAL trough characterized by three major BAL subflows (see right panels of Fig.~\ref{J123820_HeI_BAL}). However, the multi-epoch  optical spectra indicate that the corresponding \hei\ $\lambda3889$ absorption is detected only for the BAL subflow at $v_{\rm LOS}\sim-6000~\kms$ and remains generally unchanged over the epochs, characteristic of an absorber with a partial covering factor in \hei\ and being more saturated than other subflows across the entire \hei\ BAL trough.   Nevertheless, both the \hei\ and \mgii\ troughs must be saturated to some extent, because   theoretically the  optical-depth ratio between He~I* $\lambda$10830 and $\lambda$3889 for optically-thin gas is 23.3  (\citealp{Leighly11}), whereas this ratio   in the spectra of  J1238 is only $\sim$6.5 from the velocity range where the $\lambda$3889 feature is seen,  according to the apparent optical depth $\tau_a(v)= {\rm ln}[I_0(v)/I_{\rm obs} (v)]$, where $I_0(v)$ and $I_{\rm obs}(v)$ are the intrinsic and observed fluxes at velocity $v$, respectively.

\section{Discussion  } \label{discussion_sec}

\subsection{ Implications from joint analyses  } \label{origin_accel}

The four quasars, especially from group-1, are promising  candidates of BAL acceleration based on our analyses,  although other possibilities may    contribute secondarily  to the  BAL-velocity shift.   They will be treated as actual BAL-acceleration quasars throughout the following sections  to explore some of the key questions of  quasar feedback, such as the driving force of BAL acceleration, the process of BAL winds coupling  to ambient medium, and potentially observable  imprints  during the BAL-acceleration phase. Acceleration is the simplest and most  straightforward interpretation for a monolithic velocity shift of a BAL seen from multi-epoch spectra,  i.e., one would expect to see  increasing/decreasing portions of the BAL blue/red wings in acceleration cases  or   decreasing/increasing portions of the BAL blue/red wings in deceleration cases.   
Below we   explore  the  implications of BAL acceleration and then deceleration  in conjunction with the multi-wavelength data at hand.

\subsubsection{The driving mechanisms of BAL acceleration} \label{discussion_accel_mechanism}

One of the long-standing questions about  BAL winds  is the driving mechanism. Theoretically, BAL winds with LOS velocities beyond 5000 $\kms$ are thought to be driven  by  UV radiation pressure on ionized gas  within  $R\lesssim$1 pc from the quasar center,   when noticing that the typical SED of normal quasars peak at the UV band, despite some debates  about the role of  shielding  gas (e.g., \citealp{Murray95,Proga00,Hamann19}). This scenario is supported by observations and appears particularly true for LoBAL quasars, in that they are more likely X-ray weak and tend to possess softer ionizing SEDs than normal quasars (e.g., \citealp{Trump06,Gallagher06,Hamann19}).

In addition to UV radiation-driven winds on small  scales, IR radiation pressure on dust, e.g., via IR photon trapping and multi-scattering processes, may provide an additional force to accelerate ionized outflows on large scales, particularly when they are mildly optically thick to IR radiation and effectively coupled to dust (e.g., \citealp{Costa18}).  Other driving mechanisms, such as magnetic fields or cosmic rays,  may be also at work or coexist with the above two forces for accelerating BAL winds, but we are unable to assess these possibilities using the current data.  Therefore, throughout the discussion we will  ignore them and focus only on the UV/IR radiation driven scenarios.

Unlike high-velocity, UV-radiation-driven disk winds that are thought to be launched at $R\sim0.01$ pc, IR-radiation-driven outflows must exist on large-scale regions  ($R\gtrsim1$ pc)  at which dust can survive; hence, relatively low-velocity ($<$ 3000 $\kms$) outflows and  small acceleration magnitudes may be  expected if driven solely by IR radiation (e.g, \citealp{Roth12,Faucher12,Costa18}). Recently,   \citet{HeZ22} found from a small sample that  BAL velocities appear to increase  with  galactocentric distances,  from  which they interpreted as  UV-radiation-pressure on dust as the driving force of BAL acceleration,  despite the lack of investigations in kinematic  shift  as  adopted routinely in previous studies for acceleration  and   the unknown  coupling efficiency between dust and gas in that work.  Nevertheless,  IR radiation pressure on dust can exert an additional force to a high-velocity BAL wind that  is located at  relatively large radii  (e.g., $R>$ 1 pc),  leading to mild BAL acceleration as seen in group-2 quasars.

Although the IR-radiation-driven scenario offers a possible explanation  for the rarity  of  BAL acceleration as reported in \citet{Grier16}, it appears   difficult   to explain the large BAL-acceleration magnitudes  seen in the group-1 quasars.  While the large BAL acceleration from group-1 is likely  driven by UV radiation pressure at relatively small radii (e.g., $R<$ 1 pc),  the dust responsible for UV extinction/suppression may not be  necessarily  related to the presumably  traditional torus with $R<$ 10 pc; instead, dust  could  reside  in a broad range of  regions  (e.g., \citealp{Hamann17,Temple19,Rivera21}).  Thus,  whether a  BAL wind, which is launched from its accretion disk,  has reached a circumnuclear region or beyond  is crucial for the discussion.  A combined analysis of the  BAL, BEL, and continuum time-variability behaviors can  provide valuable  diagnostics, which are discussed below.

One of the most striking differences between the two group-1 quasars is the opposite time-variability pattern in UV continuum flux, such that J0136/J1344 become brighter/fainter in later epochs. If UV radiation is the dominant driver for the observed BAL acceleration, one may expect to see a  strong correlation between the continuum flux and BAL-velocity shift.  Indeed,   the largest velocity shift (acceleration) occurs in the time interval along with a decrease of UV continuum  for both quasars (see Fig.~\ref{J1344MgII_aliii_normspec} and Fig.~\ref{J013656AliiiNormFlux});  thus, the two quantities may always exhibit an opposite time-variability pattern  during an acceleration phase  or the variability in  UV continuum  may be only loosely coupled to the variability in the incident ionizing continuum seen by the BAL gas. In contrast,  the two quantities  appear to display  an opposite  variability behavior for J1259  in the interval  from  MJD=55983 to 58216  but a similar variability  behavior  in the interval from MJD=58216 to 58601. On the other hand, a much redder color  in group-2 than in group-1, again,  supports a different acceleration mechanism, such as  radiation pressure on dust  from a  large-scale region.

For the group-2 quasar J1259, \citet{ShiX16} inferred from  the two SDSS spectra that a BAL absorber is likely located at a distance of $R\sim$1 pc, on the basis of transverse motion as the cause of its BAL variability. This is also another reason for  our  caution  in identifying any  BAL acceleration based on only one spectroscopic pair.   Suppose the same BAL absorber is persistent after the second epoch, one can see from the MJD vs. Mag panel of Fig.~\ref{J1259spec6_h9_hg_hb} that the ZTF-$g$ band  monotonically brightens  while the ZTF-$r$ band remains generally unchanged over the last four spectroscopic epochs. This result, at first glance, appears to be caused by the weakening \hg\ trough after MJD=58216; however, it is difficult to explain the lack  of significant  variability in  H9 (Fig.~\ref{J1259spec6_h9_hg_hb}). We suspect that  it may signal a substantial decrease of dust or a change of dust  distribution/composition  along our LOS,  because   it is possible that the kinematic signatures of BAL acceleration/deceleration  in J1259  trace the coupling process between dust and gas, i.e.,  radiation pressure on  dust as a driver  for launching outflows and interacting with outer ISM.  For the other group-2 quasar J1238, the ZTF-$g/r$ bands show a similar light curve characterized by an  increasing  brightness over the last two spectroscopic epochs.

Interestingly, J1238 and Mrk 231 have many common features,  such as the spectral shape,   strong LoBAL (\mgii)  and weak HiBAL (\hei) troughs,  strong \feii\ and weak \oiii\ emission,  potentially inner decelerated winds and outer accelerated outflows (the decreased portion of the BAL subflow at $-12500<v_{\rm LOS}<-10000~\kms$ is comparable to the  increased portion of the BAL subflow at $-7000<v_{\rm LOS}<-5000~\kms$ from MJD=54234 to 56035; see Fig.~\ref{J123820_HeI_BAL}), and moderately weak radio emission. Therefore, like   Mrk 231 having  BAL distances at $R\sim$2--100 pc, BAL distance in J1238  may also  cover a wide  range   (see \citealp{Leighly14,Veilleux16}). If this speculation is true,  the mild acceleration seen in J1238 is also likely, at least  partly, due to IR radiation pressure on dust, although we cannot exclude  the possibility that UV radiation on ionized gas is  the dominant driver for the initial  BAL acceleration  in both quasars, particularly in cases where the dust and gas are not efficiently coupled.

 For the  group-1 quasar J1344,  an  intriguing result is  the progressively  broadening   BAL-2 as opposed to  narrowing  BAL-1, such that the width  increment of BAL-2  is  almost  equal to the width decrement of BAL-1  from MJD=53503 to 59384 in both \aliii\ and \mgii\ (see the MJD vs. $w_{95}$ panel in Fig.~\ref{J1344MgII_aliii_normspec}),  possibly indicative of a new BAL-acceleration event characterized by a high-velocity inner wind (BAL-1  subflow) overtaking  a low-velocity outer wind.   This scenario is supported  by  the disappearance of both the  \aliii\  BAL (at $v_{\rm LOS}\sim-11000 \kms$) and the \aliii\ BEL peak, as well as the rapid  drop of the \civ\ BEL  strength after MJD = 58224,  an epoch when the BAL-1 started to level off or perhaps decelerate.    For the other group-1 quasar J0136, the disappearance of LoBAL ions that are  associated with a  decrease of dust, could be the main cause for the brightening UV continuum;  hence, its largest BAL acceleration magnitude  detected between MJD=52203 and 55444,  could be dominated by UV-radiation-pressure on dusty gas.  This interpretation is further  supported by a decline of the acceleration magnitude  after MJD=55444 (see Fig.~\ref{J013656AliiiNormFlux}), an epoch when  LoBALs  and dust started  to  disappear  along our LOS.   It is worth noting that  the  time-resolved evidence  for large-magnitude BAL acceleration along with line-locking,  associated absorption lines (AALs; see \citealp{Weymann91} and references therein)  seen in group-1 quasars,  offers a plausible interpretation  to link  nuclear BAL winds and galactic AAL outflows, despite the huge difference in distance ($<1$ pc vs. $>$1,000 pc).   As a comparison,  spatially resolved evidence  for small-magnitude  acceleration/deceleration  traced  by the positive/negative correlations between \oiii\ emission-line width and  velocity   has  been  reported  in the literature (e.g., FWHM increasing due to turbulence from acceleration or deceleration;  see  Fig.~5 in \citealt{Nesvadba06}). Therefore,  spatially resolved spectroscopy of the BAL-acceleration quasars in the future  may provide  unique and valuable  information  to bridge the huge gap of distance.

A joint analysis of variability in both absorption and emission may offer additional  diagnostics  to further advance our understanding of BAL acceleration.  It is clear that the BALs and BELs vary dramatically in J1344, while the \mgii/\aliii\ BELs in J0136 remain generally unchanged as opposed to complete disappearance of  the  \mgii/\aliii\ BALs. In particular, the \aliii\  BEL peak  in J1344  exhibited a  velocity shift comparable  to that of the \aliii\ BAL-1 in the interval from MJD=53503  to 58224,  despite the  different  acceleration  magnitudes measured in between. Such a difference can be   explained by  the  complex  structures and inner physics of the  BEL outflows.  Moreover, the \aliii\ BEL in J1344 became  progressively  weak  after  MJD=56363  (its peak nearly disappeared at  MJD=59384; see  Fig.~\ref{J1344MgII_aliii_normspec}),  again,  supportive of UV-radiation pressure  as a common driving force to  accelerate  both the  BAL-1 and \aliii\ BEL outflows. Given the  brightening/dimming UV continuum detected in J0136/J1344,   we argue  that the absorption/emission-line  variability behaviors,  such as the monolithic BAL shifts   and  large REW drops of the \civ\ BEL,  are loosely linked to UV continuum variability.

In addition,  the BAL/BEL signal the LOS/bulk effects, respectively, which may also  contribute  to  the  differences of variability behaviors  between J0136 and J1344.  However, it remains difficult to explain (1) the opposite  time-variability pattern in REW between BAL-1 and BAL-2  for J1344, and (2) the rapidly  disappeared \aliii\ BEL as opposed to the generally unchanged \mgii\ BEL  for  J1344 after MJD=58224. The second phenomenon may be  linked to the third \aliii\ BAL at $v_{\rm LOS}\sim-11000~\kms$, whose  variability behavior cannot be  assessed in detail  given its  presence only at MJD=56363.   Interestingly, both quasars show a common variability pattern that the \civ\ BEL strength  decreased by a factor of $\sim$2 near  the point where the BAL velocity starts to level off, again, supporting a connection between the BAL and BEL outflows; furthermore, both quasars exhibit slightly blueshifted, line-locking \civ\ signatures,  indicative of  quasar  radiation-driven outflows  coupled to  large-scale ISM.  

It is worth noting again that the smaller the BAL-velocity shift between two epochs, the greater the chance that the shift could be due to velocity-dependent variability caused by ionization changes and/or transverse motions. Therefore, it is possible  that another agent comes into play at some point in term of curtailing acceleration, particularly when noticing the plateau-like trend among them in later epochs (see Fig.~\ref{cut_velocity_shift_vs_dt_Grier16}).

\subsubsection{ BAL deceleration as an origin of weak radio emission and large-scale AALs} \label{decel_radio}

The presence of radio emission in BAL quasars  remains unknown  since BAL winds  are thought  to be launched from a viewing angle closer to the equatorial plane of  accretion disk than the jet (e.g., \citealp{Becker2000,Yi19a,Nair2022}). However, 
both theoretical and observational studies of the wind-ISM interactions  predict  the production  of weak radio emission (e.g., \citealp{Faucher12,Zakamska14}), which were indeed detected in the group-2 quasars (J1238 with a flux density of 2.5/5.84 mJy and J1259 with a flux density of 0.8/2.0 mJy at 3.0/1.4 GHz, respectively).   Interestingly, the  two radio-band observations yield a  spectrum index of $\alpha \sim-1.1$ for both quasars,  reinforcing  the argument for the wind-ISM interaction as an origin of weak radio emission (\citealp{Panessa19}).  Moreover, all but J1344  were also detected by RACS at $\sim$800 MHz ($\sim$8.3/2.2/1.7 mJy for J1238/J1259/J0136, respectively, while out  of the survey range for J1344; \citealp{McConnell2020}), consistent with a higher  radio-detection rate at a lower frequency  in the LoBAL population (\citealt{Morabito19}).   Given  that the LoBAL  quasar Mrk 231 also  has moderately weak radio emission along with evidence for BAL deceleration/acceleration and radio flares,  we   propose  an underlying link between the two phenomena, which can be tested with follow-up observations in the future.

As a comparison, \citet{ShiX16}  speculated that  the Balmer absorber in J1259  experienced deceleration due to the collision with  surrounding medium,   based on an analysis of the  two SDSS spectra.  In combination with  the observational results  from  six optical spectra and multi-wavelength data,  we did find  additional  evidence in support of  wind  deceleration  for J1259.  Furthermore,  such  deceleration  events  can explain  the high incidence of weak radio emission  among red quasars that  appears to peak  around the radio-quiet and radio-loud threshold (\citealp{Klindt19}),  given an  anomalously  high fraction of BALs  seen  in  red quasars (e.g., \citealp{Urrutia09,Fynbo13,Hamann17}) and  a high radio-detection rate  found   in the BAL population (e.g., \citealp{Yi19a,Morabito19}).

Like the group-1 quasars,  three X-ray bright BAL quasars from  \citet{Joshi14,Joshi19} also possess AALs and kinematic signatures of BAL deceleration.   The lack of significant velocity shifts in AAL   as opposed to large BAL-velocity shifts found both in  this work and \citet{Joshi14,Joshi19}, suggests that  large-magnitude, identifiable  acceleration events mostly occur  in BALs rather than AALs,  consistent with the latter being located at a large-scale  region, i.e.,  the quasar host galaxy or a cluster near the quasar (e.g., \citealp{Weymann91}); furthermore, a complex AAL without significant variability over decades may  trace  an outermost  remnant of  BAL winds fully coupled to ISM, given the  presence of wide-spread,  line-locking signatures  in  the group-1 quasars.  
The group-2  quasars, however, have lower systemic redshifts than group-1  quasars,  making an identification of   \civ\ AALs impossible from  optical spectroscopy.    These BAL-acceleration candidates  lack or  possess  shallow  \mgii\  AALs,  indicating that AALs tend to be in  a relatively high-ionization state,  although  it is difficult to  identify  shallow  \mgii\ AAL features from group-2  due  to  overlapping absorption.   
Additional insights of BAL-acceleration/deceleration and their effects may be gained  from  X-ray  observations,  particularly in the context that BAL wind  is thought to be associated with the shielding gas.

\subsection{ Implications from  quasar  color      } \label{radio_red}   
All  four quasars have a SED shape consistent with that of red quasars (see \citealt{Rivera21});  however, the group-2 quasars exhibit weak radio emission and are much redder than the group-1 quasars (see  Fig.~\ref{J1344sed_2epoch}).  If radio emission in group-2 signals a small viewing angle to the jet axis and the group-1 LOS is  seen through  the edge of a traditional  torus, then the group-2 quasars are expected to be bluer than group-1, which is opposite to the observations.  As discussed above, the weak radio emission in group-2 is likely a byproduct from BAL-ISM interactions rather than a tracer of  low-power  jets.

Here, we explore whether the redder color in group-2 is also a consequence of the BAL-ISM interaction.  In stark contrast with the rest-frame UV continuum variability detected in the four quasars, none  showed  significant variability in  the W1 and W2 bands. This difference, along with the partial LOS covering for BAL winds,  suggests that the UV continuum variability is likely due to a patchy effect  caused by rapid  changes  in dust composition, distribution, and/or transverse motions  during the  BAL-acceleration phase.   Indeed,  the   dimming   $V$-band  brightness in J1259  (see Fig.~\ref{J1259spec6_h9_hg_hb}) is   at least  partly  caused  by the BAL-ISM  coupling,  a process that can produce  rapid changes of the dust/gas in distribution and composition, making  in situ dust formation and patchy obscuration possible.

All the four quasars have a SED bump at $\lambda_{\rm rest}\sim3~\mu$m, in  agreement  with the prediction of dusty winds (e.g., \citealp{ZhangS14,Gallagher15,Rivera21}). Importantly, the group-2 quasars have a steeper SED rise  at $1<\lambda_{\rm rest}<3~\mu$m   than  group-1 quasars, providing  further  evidence for  the correlation between near-IR slope and BAL properties  as reported  in \citet{ZhangS14}, where they speculated that BAL winds are strongly decelerated by interacting  with  ISM.  Our observations  suggest  that  BAL deceleration may play a more important role than BAL acceleration with respect to controlling the SED shape, perhaps due to  stronger in situ dust formation  and/or  entrainments  from  an  inhomogeneous/clumpy/patchy environment.  This result  is  intriguing and sheds  light on the nature of red and blue quasars, especially those with  outflows  undergoing  actual acceleration, as the W3/SDSS-$i$ ratio may change dramatically on short  timescales due to changes solely by dust variability along our LOS.  Likewise, the conventional classification of  radio-loud and radio-quiet quasars would be problematic  in such cases (see an alternative tracer of radio loudness proposed  by  \citealp{Klindt19}).   However, the relations between BAL acceleration and UV  reddening or brightness are unclear, given  that  J0136/J1259 become brighter/dimmer  in the UV band while both show an increase in UV reddening over the epochs, and that  J1238/J1344 become brighter/dimmer in the UV band while both remain generally unchanged in UV reddening. 
Apart from the inhomogeneous/clumpy/patchy environment,  different dust variability behaviors at different wavelengths may be also related to dust composition  or in situ  dust formation, i.e., the wind-ISM interaction can break dense clouds into diffuse filaments, making more dust exposed to quasar UV radiation  and hence enhancing IR emission  (e.g., \citealp{Wagner13,Hamann17}).

\section{Summary}

In this work, we select from our sample four  LoBAL-acceleration candidates and investigate their physical properties based on multi-wavelength, multi-epoch observations,  aiming to  bridge the gap over HiBAL acceleration and gain unique  insights into actual quasar feedback.  The main observational results and conclusions are summarized below.

\begin{enumerate}
\item  We identified one compelling (J1344), one strong (J0136), and two tentative  cases of LoBAL acceleration, among which J0136 exhibited  BAL  disappearance in \aliii\ and \mgii\ (see Section.~\ref{accel_identification}).   

\item
The group-1 (J1344 and J0136) are  radio quiet  and exhibit line-locking signatures in \civ\ AALs, while the group-2 (J1238 and J1259) have  red SEDs  and weak radio emission with a steep  spectral index ($\alpha\sim -1.1$) (see Section.~\ref{sec_SED} and \ref{decel_radio}).  

\item
The BAL-ISM  coupling  is  one of the major avenues for the origin of quasar reddening, patchy obscuration, large-scale AALs, and perhaps weak  radio emission (see Section.~\ref{radio_red}). 

\end{enumerate}

All the four quasars exhibit BAL-acceleration magnitudes larger than the detection upper limits of BAL acceleration derived from \citet{Grier16},  which deserve multi-wavelength, follow-up observations, particularly J1259 with  one-sided,  off-nuclear \oii\ emission nebula.  They  may  provide  an ideal laboratory  to  study  the actual feedback processes,  such as   strong  deceleration of BAL winds  before traveling  out to large scales,  the  process  of  BAL winds  breaking out  a circumnuclear  dust cocoon (e.g., \citealp{ZhangS14,ShiX16,Temple19}),  the origin of giant nebulae from a non-galaxy-cluster environment,  the relation between  mergers and star-formation/quasar activities,  and   gravitational-wave recoiling SMBHs  etc,  in the context that LoBALs  are representative of young quasars   from  gas-rich mergers  (e.g., \citealp{Urrutia09,Yi22}).

\section{acknowledgments}

We thank the anonymous referee for a thorough review and constructive comments which led to an improved manuscript. 
We thank for Pu Du, Chen Hu, and Minfeng Gu for stimulating discussions.  
We are grateful to Feige Wang, Jingyi Yang, and Xiaohui Fan for assistance with the LBT/MODS observations.   
W.Y.  thanks support from the National Science Foundation of China (NSFC; 11703076).  P.B.H. is supported by NSERC grant 2023-05068. Z.Y. is supported by NSFC (12073069) and the Xiaoxiang Scholars Programme of Hunan Normal University. Zhicheng He acknowledges support from NSFC (12222304 and 12192221).  W.Y.,  J.M.B, and X.B.W. acknowledge the science research grants from the China Manned Space Project with No. CMS-CSST-2021-A06.

This research  uses data  obtained through the Telescope Access Program (TAP), which has been funded by the National Astronomical Observatories of China, the Chinese Academy of Sciences (the Strategic Priority Research Program ''The Emergence of Cosmological Structures'' grant No. XDB09000000), and the Special Fund for Astronomy from the Ministry of Finance. Observations obtained with the Hale Telescope at Palomar Observatory were obtained as part of an agreement between the National Astronomical Observatories, the Chinese Academy of Sciences, and the California Institute of Technology. 
The Hobby-Eberly Telescope (HET) is a joint project of the University of Texas at Austin, the Pennsylvania State University, Ludwig-Maximillians-Universit\"{a}t M\"{u}nchen, and Georg-August-Universit\"{a}t G\"{o}ttingen. The Hobby-Eberly Telescope is named in honour of its principal benefactors, William P. Hobby and Robert E. Eberly. 
The Low-Resolution Spectrograph 2 (LRS2) was developed and funded by the
University of Texas at Austin McDonald Observatory and Department of
Astronomy, and by the Pennsylvania State University. We thank the
Leibniz-Institut f\"ur Astrophysik Potsdam and the Institut f\"ur
Astrophysik G\"ottingen for their contributions to the construction
of the integral field units. 
The LBT is an international collaboration among institutions in the United States, Italy and Germany. LBT Corporation partners are: The University of Arizona on behalf of the Arizona Board of Regents; Istituto Nazionale di Astrofisica, Italy; LBT Beteiligungsgesellschaft, Germany, representing the Max-Planck Society, The Leibniz Institute for Astrophysics Potsdam, and Heidelberg University; The Ohio State University, representing OSU, University of Notre Dame, University of Minnesota and University of Virginia.  
We acknowledge the support of the staff of the Lijiang 2.4 m telescope (LJT). Funding for the telescope has been provided by CAS and the People’s Government of Yunnan Province.


\begin{thebibliography}{}
\small

\bibitem[\protect\citeauthoryear{Arav et al.}{2015}]{Arav15} Arav N., Chamberlain C., Kriss G.~A., Kaastra J.~S., Cappi M., Mehdipour M., Petrucci P.-O., et al., 2015, A\&A, 577, A37. doi:10.1051/0004-6361/201425302
\bibitem[Arav et al.(2018)]{Arav18}Arav, N., Liu, G., Xu, X., et al. 2018, \apj, 857, 60 
\bibitem[\protect\citeauthoryear{Aromal, Srianand, \& Petitjean}{2021}]{Aromal21} Aromal P., Srianand R., Petitjean P., 2021, MNRAS, 504, 5975. doi:10.1093/mnras/stab1299
\bibitem[Becker et al.(2000)]{Becker2000} Becker, R.~H., White, R.~L., Gregg, M.~D., et al.\ 2000, \apj, 538, 72. doi:10.1086/309099
\bibitem[\protect\citeauthoryear{Blanton et al.}{2017}]{Blanton17} Blanton M.~R., Bershady M.~A., Abolfathi B., Albareti F.~D., Allende Prieto C., Almeida A., Alonso-Garc{\'\i}a J., et al., 2017, AJ, 154, 28. doi:10.3847/1538-3881/aa7567
\bibitem[\protect\citeauthoryear{Byun, Arav, \& Hall}{2022}]{Byun22} Byun D., Arav N., Hall P.~B., 2022, ApJ, 927, 176. doi:10.3847/1538-4357/ac503d
\bibitem[\protect\citeauthoryear{Calistro Rivera et al.}{2021}]{Rivera21} Calistro Rivera G., Alexander D.~M., Rosario D.~J., Harrison C.~M., Stalevski M., Rakshit S., Fawcett V.~A., et al., 2021, A\&A, 649, A102. doi:10.1051/0004-6361/202040214
\bibitem[\protect\citeauthoryear{Capellupo et al.}{2011}]{Capellupo11} Capellupo D.~M., Hamann F., Shields J.~C., Rodr{\'\i}guez Hidalgo P., Barlow T.~A., 2011, MNRAS, 413, 908. doi:10.1111/j.1365-2966.2010.18185.x
\bibitem[\protect\citeauthoryear{Chonis et al.}{2014}]{Chonis14} Chonis T.~S., Hill G.~J., Lee H., Tuttle S.~E., Vattiat B.~L., 2014, SPIE, 9147, 91470A. doi:10.1117/12.2056005
\bibitem[Davis et al.(2018)]{Davis18} Davis, B.~D., Ciardullo, R., Jacoby, G.~H., et al.\ 2018, \apj, 863, 189
\bibitem[\protect\citeauthoryear{Costa et al.}{2018}]{Costa18} Costa T., Rosdahl J., Sijacki D., Haehnelt M.~G., 2018, MNRAS, 473, 4197. doi:10.1093/mnras/stx2598
\bibitem[\protect\citeauthoryear{Eisenstein et al.}{2011}]{Eisenstein11} Eisenstein D.~J., Weinberg D.~H., Agol E., Aihara H., Allende Prieto C., Anderson S.~F., Arns J.~A., et al., 2011, AJ, 142, 72. doi:10.1088/0004-6256/142/3/72
\bibitem[\protect\citeauthoryear{Fabian}{2012}]{Fabian12} Fabian A.~C., 2012, ARA\&A, 50, 455. doi:10.1146/annurev-astro-081811-125521
\bibitem[\protect\citeauthoryear{Fan et al.}{2015}]{FanY15} Fan Y.-F., Bai J.-M., Zhang J.-J., Wang C.-J., Chang L., Xin Y.-X., Zhang R.-L., 2015, RAA, 15, 918. doi:10.1088/1674-4527/15/6/014
\bibitem[\protect\citeauthoryear{Faucher-Gigu{\`e}re \& Quataert}{2012}]{Faucher12} Faucher-Gigu{\`e}re C.-A., Quataert E., 2012, MNRAS, 425, 605. doi:10.1111/j.1365-2966.2012.21512.x
\bibitem[\protect\citeauthoryear{Filiz Ak et al.}{2013}]{Filizak13} Filiz Ak N., Brandt W.~N., Hall P.~B., Schneider D.~P., Anderson S.~F., Hamann F., Lundgren B.~F., et al., 2013, ApJ, 777, 168. doi:10.1088/0004-637X/777/2/168
\bibitem[\protect\citeauthoryear{Fynbo et al.}{2013}]{Fynbo13} Fynbo J.~P.~U., Krogager J.-K., Venemans B., Noterdaeme P., Vestergaard M., M{\o}ller P., Ledoux C., et al., 2013, ApJS, 204, 6. doi:10.1088/0067-0049/204/1/6
\bibitem[\protect\citeauthoryear{Fu \& Stockton}{2009}]{FuH09} Fu H., Stockton A., 2009, ApJ, 690, 953. doi:10.1088/0004-637X/690/1/953
\bibitem[\protect\citeauthoryear{Gabel et al.}{2003}]{Gabel03} Gabel J.~R., Crenshaw D.~M., Kraemer S.~B., Brandt W.~N., George I.~M., Hamann F.~W., Kaiser M.~E., et al., 2003, ApJ, 595, 120. doi:10.1086/377342
\bibitem[\protect\citeauthoryear{Gallagher et al.}{2006}]{Gallagher06} Gallagher S.~C., Brandt W.~N., Chartas G., Priddey R., Garmire G.~P., Sambruna R.~M., 2006, ApJ, 644, 709. doi:10.1086/503762
\bibitem[\protect\citeauthoryear{Gallagher et al.}{2015}]{Gallagher15} Gallagher S.~C., Everett J.~E., Abado M.~M., Keating S.~K., 2015, MNRAS, 451, 2991. doi:10.1093/mnras/stv1126
\bibitem[Grier et al.(2016)]{Grier16} Grier, C.~J., Brandt, W.~N., Hall, P.~B., et al.\ 2016, \apj, 824, 130
\bibitem[\protect\citeauthoryear{Hall et al.}{2007a}]{Hall07a} Hall P.~B., Sadavoy S.~I., Hutsemekers D., Everett J.~E., Rafiee A., 2007a, ApJ, 665, 174. doi:10.1086/519273
\bibitem[\protect\citeauthoryear{Hall}{2007b}]{Hall07b} Hall P.~B., 2007b, AJ, 133, 1271. doi:10.1086/511272
\bibitem[\protect\citeauthoryear{Hamann et al.}{2017}]{Hamann17} Hamann F., Zakamska N.~L., Ross N., Paris I., Alexandroff R.~M., Villforth C., Richards G.~T., et al., 2017, MNRAS, 464, 3431. doi:10.1093/mnras/stw2387
\bibitem[\protect\citeauthoryear{Hamann et al.}{2019}]{Hamann19} Hamann F., Herbst H., Paris I., Capellupo D., 2019, MNRAS, 483, 1808. doi:10.1093/mnras/sty2900
\bibitem[\protect\citeauthoryear{Hemler et al.}{2019}]{Helmer19} Hemler Z.~S., Grier C.~J., Brandt W.~N., Hall P.~B., Horne K., Shen Y., Trump J.~R., et al., 2019, ApJ, 872, 21. doi:10.3847/1538-4357/aaf1bf
\bibitem[\protect\citeauthoryear{He et al.}{2019}]{HeZ19} He Z., Wang T., Liu G., Wang H., Bian W., Tchernyshyov K., Mou G., et al., 2019, NatAs, 3, 265. doi:10.1038/s41550-018-0669-8
\bibitem[\protect\citeauthoryear{He et al.}{2022}]{HeZ22} He Z., Liu G., Wang T., Mou G., Green R., Bian W., Wang H., et al., 2022, SciA, 8, eabk3291. doi:10.1126/sciadv.abk3291
\bibitem[\protect\citeauthoryear{Hill et al.}{2021}]{Hill21} Hill G.~J., Lee H., MacQueen P.~J., Kelz A., Drory N., Vattiat B.~L., Good J.~M., et al., 2021, AJ, 162, 298. doi:10.3847/1538-3881/ac2c02
\bibitem[\protect\citeauthoryear{Jester et al.}{2005}]{Jester2005} Jester S., Schneider D.~P., Richards G.~T., Green R.~F., Schmidt M., Hall P.~B., Strauss M.~A., et al., 2005, AJ, 130, 873. doi:10.1086/432466
\bibitem[\protect\citeauthoryear{Joshi et al.}{2014}]{Joshi14} Joshi R., Chand H., Srianand R., Majumdar J., 2014, MNRAS, 442, 862. doi:10.1093/mnras/stu840
\bibitem[\protect\citeauthoryear{Joshi et al.}{2019}]{Joshi19} Joshi R., Srianand R., Chand H., Wu X.-B., Noterdaeme P., Petitjean P., Ho L.~C., 2019, ApJ, 871, 43. doi:10.3847/1538-4357/aaf500
\bibitem[\protect\citeauthoryear{Kriss et al.}{2019}]{Kriss19} Kriss G.~A., Mehdipour M., Kaastra J.~S., Rau A., Bodensteiner J., Plesha R., Arav N., et al., 2019, A\&A, 621, A12. doi:10.1051/0004-6361/201834326
\bibitem[\protect\citeauthoryear{Klindt et al.}{2019}]{Klindt19} Klindt L., Alexander D.~M., Rosario D.~J., Lusso E., Fotopoulou S., 2019, MNRAS, 488, 3109. doi:10.1093/mnras/stz1771
\bibitem[\protect\citeauthoryear{Leighly, Dietrich, \& Barber}{2011}]{Leighly11} Leighly K.~M., Dietrich M., Barber S., 2011, ApJ, 728, 94. doi:10.1088/0004-637X/728/2/94
\bibitem[\protect\citeauthoryear{Leighly et al.}{2014}]{Leighly14} Leighly K.~M., Terndrup D.~M., Baron E., Lucy A.~B., Dietrich M., Gallagher S.~C., 2014, ApJ, 788, 123. doi:10.1088/0004-637X/788/2/123
\bibitem[\protect\citeauthoryear{Lyke et al.}{2020}]{Lyke20} Lyke B.~W., Higley A.~N., McLane J.~N., Schurhammer D.~P., Myers A.~D., Ross A.~J., Dawson K., et al., 2020, ApJS, 250, 8. doi:10.3847/1538-4365/aba623
\bibitem[\protect\citeauthoryear{Lu \& Lin}{2020}]{LuW20} Lu W.-J., Lin Y.-R., 2020, MNRAS, 499, L58. doi:10.1093/mnrasl/slaa158
\bibitem[\protect\citeauthoryear{McGraw et al.}{2017}]{McGraw17} McGraw S.~M., Brandt W.~N., Grier C.~J., Filiz Ak N., Hall P.~B., Schneider D.~P., Anderson S.~F., et al., 2017, MNRAS, 469, 3163. doi:10.1093/mnras/stx1063
\bibitem[McConnell et al.(2020)]{McConnell2020} McConnell, D., Hale, C.~L., Lenc, E., et al.\ 2020, PASA, 37, e048. doi:10.1017/pasa.2020.41
\bibitem[\protect\citeauthoryear{Morabito et al.}{2019}]{Morabito19} Morabito L.~K., Matthews J.~H., Best P.~N., G{\"u}rkan G., Jarvis M.~J., Prandoni I., Duncan K.~J., et al., 2019, A\&A, 622, A15. doi:10.1051/0004-6361/201833821
\bibitem[\protect\citeauthoryear{Murray et al.}{1995}]{Murray95} Murray N., Chiang J., Grossman S.~A., Voit G.~M., 1995, ApJ, 451, 498. doi:10.1086/176238
\bibitem[Nair \& Vivek(2022)]{Nair2022} Nair, A. \& Vivek, M.\ 2022, \mnras, 511, 4946. doi:10.1093/mnras/stac204
\bibitem[\protect\citeauthoryear{Nesvadba et al.}{2006}]{Nesvadba06} Nesvadba N.~P.~H., Lehnert M.~D., Eisenhauer F., Gilbert A., Tecza M., Abuter R., 2006, ApJ, 650, 693. doi:10.1086/507266
\bibitem[\protect\citeauthoryear{Panessa et al.}{2019}]{Panessa19} Panessa F., Baldi R.~D., Laor A., Padovani P., Behar E., McHardy I., 2019, NatAs, 3, 387. doi:10.1038/s41550-019-0765-4
\bibitem[Proga et al.(2000)]{Proga00} Proga, D., Stone, J.~M., \& Kallman, T.~R.\ 2000, \apj, 543, 686 
\bibitem[Rogerson et al.(2016)]{Rogerson16} Rogerson, J.~A., Hall, P.~B., Rodr{\'\i}guez Hidalgo, P., et al.\ 2016, \mnras, 457, 405
\bibitem[Rogerson et al.(2018)]{Rogerson18} Rogerson, J.~A., Hall, P.~B., Ahmed, N.~S., et al.\ 2018, \apj, 862, 22 
\bibitem[Rodr{\'\i}guez Hidalgo et al.(2020)]{Paola20} Rodr{\'\i}guez Hidalgo, P., Khatri, A.~M., Hall, P.~B., et al.\ 2020, \apj, 896, 151    
\bibitem[\protect\citeauthoryear{Roth et al.}{2012}]{Roth12} Roth N., Kasen D., Hopkins P.~F., Quataert E., 2012, ApJ, 759, 36. doi:10.1088/0004-637X/759/1/36
\bibitem[Rafiee et al.(2016)]{Rafiee16} Rafiee, A., Pirkola, P., Hall, P.~B., et al.\ 2016, \mnras, 459, 2472  
\bibitem[Ramsey et al.(1998)]{Ramsey98} Ramsey, L.~W., Adams, M.~T., Barnes, T.~G., et al.\ 1998, \procspie, 3352, 34
\bibitem[Rankine et al.(2020)]{Rankine20} Rankine, A.~L., Hewett, P.~C., Banerji, M., et al.\ 2020, \mnras, 492, 4553
\bibitem[\protect\citeauthoryear{Reynolds et al.}{2013}]{Reynolds13} Reynolds C., Punsly B., O'Dea C.~P., Hurley-Walker N., 2013, ApJL, 776, L21. doi:10.1088/2041-8205/776/2/L21
\bibitem[\protect\citeauthoryear{Ross et al.}{2020}]{Ross20} Ross N.~P., Graham M.~J., Calderone G., Ford K.~E.~S., McKernan B., Stern D., 2020, MNRAS, 498, 2339. doi:10.1093/mnras/staa2415
\bibitem[\protect\citeauthoryear{Urrutia et al.}{2009}]{Urrutia09} Urrutia T., Becker R.~H., White R.~L., Glikman E., Lacy M., Hodge J., Gregg M.~D., 2009, ApJ, 698, 1095. doi:10.1088/0004-637X/698/2/1095
\bibitem[\protect\citeauthoryear{Shi et al.}{2016}]{ShiX16} Shi X., Zhou H., Shu X., Zhang S., Ji T., Pan X., Sun L., et al., 2016, ApJ, 819, 99. doi:10.3847/0004-637X/819/2/99
\bibitem[\protect\citeauthoryear{Scott et al.}{2014}]{Scott14} Scott A.~E., Brandt W.~N., Behar E., Crenshaw D.~M., Gabel J.~R., Gibson R.~R., Kaspi S., et al., 2014, ApJ, 797, 105. doi:10.1088/0004-637X/797/2/105
\bibitem[\protect\citeauthoryear{Temple et al.}{2019}]{Temple19} Temple M.~J., Banerji M., Hewett P.~C., Coatman L., Maddox N., Peroux C., 2019, MNRAS, 487, 2594. doi:10.1093/mnras/stz1420
\bibitem[\protect\citeauthoryear{Trump et al.}{2006}]{Trump06} Trump J.~R., Hall P.~B., Reichard T.~A., Richards G.~T., Schneider D.~P., Vanden Berk D.~E., Knapp G.~R., et al., 2006, ApJS, 165, 1. doi:10.1086/503834
\bibitem[\protect\citeauthoryear{Veilleux et al.}{2016}]{Veilleux16} Veilleux S., Mel{\'e}ndez M., Tripp T.~M., Hamann F., Rupke D.~S.~N., 2016, ApJ, 825, 42. doi:10.3847/0004-637X/825/1/42
\bibitem[\protect\citeauthoryear{Wang et al.}{2019}]{WangC19} Wang C.-J., Bai J.-M., Fan Y.-F., Mao J.-R., Chang L., Xin Y.-X., Zhang J.-J., et al., 2019, RAA, 19, 149. doi:10.1088/1674-4527/19/10/149
\bibitem[\protect\citeauthoryear{Wagner, Umemura, \& Bicknell}{2013}]{Wagner13} Wagner A.~Y., Umemura M., Bicknell G.~V., 2013, ApJL, 763, L18. doi:10.1088/2041-8205/763/1/L18
\bibitem[Wang et al.(2015)]{WangT15} Wang, T., Yang, C., Wang, H., et al.\ 2015, \apj, 814, 150
\bibitem[Weymann et al.(1991)]{Weymann91} Weymann, R.~J., Morris, S.~L., Foltz, C.~B., et al.\ 1991, \apj, 373, 23
\bibitem[Wilson et al.(2004)]{Wilson04} Wilson, J.~C., Henderson, C.~P., Herter, T.~L., et al.\ 2004, \procspie, 5492, 1295
\bibitem[Xu et al.(2020)]{XuX20} Xu, X., Zakamska, N.~L., Arav, N., et al.\ 2020, \mnras, 495, 305
\bibitem[Yi et al.(2019)]{Yi19a}Yi, W.; Brandt, W. N.; Hall, P. B., et al. 2019, \apjs, 242, 28
\bibitem[Yi \& Timlin(2021)]{Yi21} Yi, W. \& Timlin, J.\ 2021, \apjs, 255, 12
\bibitem[\protect\citeauthoryear{Yi et al.}{2022}]{Yi22} Yi W., Brandt W.~N., Ni Q., Ho L.~C., Luo B., Yan W., Schneider D.~P., et al., 2022, ApJ, 930, 5. doi:10.3847/1538-4357/ac6109
\bibitem[\protect\citeauthoryear{York et al.}{2000}]{York00} York D.~G., Adelman J., Anderson J.~E., Anderson S.~F., Annis J., Bahcall N.~A., Bakken J.~A., et al., 2000, AJ, 120, 1579. doi:10.1086/301513
\bibitem[\protect\citeauthoryear{Zakamska \& Greene}{2014}]{Zakamska14} Zakamska N.~L., Greene J.~E., 2014, MNRAS, 442, 784. doi:10.1093/mnras/stu842
\bibitem[\protect\citeauthoryear{Zhang et al.}{2014}]{ZhangS14} Zhang S., Wang H., Wang T., Xing F., Zhang K., Zhou H., Jiang P., 2014, ApJ, 786, 42. doi:10.1088/0004-637X/786/1/42

\end{thebibliography}
\end{document}